\documentclass[12pt]{article}
\pdfoutput=1

\usepackage{amssymb}
\usepackage{amsmath}
\usepackage{bbold}
\usepackage{color}
\usepackage{colordvi}
\usepackage{fancybox}
\usepackage[footnotesize]{caption2}
\usepackage{graphicx}
\usepackage[center,footnotesize,hang]{subfigure}
\usepackage{bbm}
\usepackage{url}
\usepackage{multirow}
\usepackage{array}
\usepackage{arydshln}
\usepackage{pifont}
\usepackage{mathrsfs}
\usepackage{fancybox}
\usepackage{cite}
\usepackage[colorlinks]{hyperref}
\usepackage{enumitem}  
\usepackage{float} %
\newcommand{\PreserveBackslash}[1]{\let\temp=\\#1\let\\=\temp}
\newcolumntype{C}[1]{>{\PreserveBackslash\centering}p{#1}}
\newcolumntype{R}[1]{>{\PreserveBackslash\raggedleft}p{#1}}
\newcolumntype{L}[1]{>{\PreserveBackslash\raggedright}p{#1}}
\newcommand{\cleqn}{\setcounter{equation}{0}}
\allowdisplaybreaks \allowdisplaybreaks[2]

\makeatletter
\@addtoreset{equation}{section}
\makeatother

\begin{document}

\title{
\begin{flushright}
\hfill\mbox{\small USTC-ICTS-17-05} \\[5mm]
\begin{minipage}{0.2\linewidth}
\normalsize
\end{minipage}
\end{flushright}
{\Large \bf Golden Littlest Seesaw
\\[2mm]}}

\date{}

\author{
Gui-Jun~Ding$^{1}$\footnote{E-mail: {\tt
dinggj@ustc.edu.cn}},  \
Stephen~F.~King$^{2}$\footnote{E-mail: {\tt king@soton.ac.uk}}, \
Cai-Chang Li$^{1}$\footnote{E-mail: {\tt
lcc0915@mail.ustc.edu.cn}}  \
\\*[20pt]
\centerline{
\begin{minipage}{\linewidth}
\begin{center}
$^1${\it \small
Interdisciplinary Center for Theoretical Study and  Department of Modern Physics,\\
University of Science and Technology of China, Hefei, Anhui 230026, China}\\[2mm]
$^2${\it \small
Physics and Astronomy,
University of Southampton,
Southampton, SO17 1BJ, U.K.}\\
\end{center}
\end{minipage}}
\\[10mm]}
\maketitle
\thispagestyle{empty}

\begin{abstract}
\noindent
We propose and analyse a new class of Littlest Seesaw models, with two right-handed neutrinos in their diagonal mass basis, based on preserving the first column of the Golden Ratio mixing matrix. We perform an exhaustive analysis of all possible remnant symmetries of the group $A_5$ which can be used to enforce various vacuum alignments for the flavon controlling solar mixing, for two simple cases of the atmospheric flavon vacuum alignment. The solar and atmospheric flavon vacuum alignments are enforced by {\em different} remnant symmetries. We examine the phenomenological viability of each of the possible Littlest Seesaw alignments in $A_5$, which preserve the first column of the Golden ratio mixing matrix, using figures and extensive tables of benchmark points and comparing our predictions to a recent global analysis of neutrino data. We also repeat the analysis for an alternative form of Golden Ratio mixing matrix.
\end{abstract}
\newpage

\section{\label{sec:introduction}Introduction}
\indent

Massive neutrinos together with neutrino oscillations has been firmly established, and it is unique experimental evidence for physics beyond the standard model. All the three lepton mixing angles $\theta_{12}$, $\theta_{13}$ and $\theta_{23}$ and the mass squared differences $\delta m^2\equiv m^2_2-m^2_1$ and $\Delta m^2\equiv m^2_3-(m^2_1+m^2_2)/2$ has been precisely measured in a large number of neutrino oscillation experiments. At present the $3\sigma$ ranges of these mixing parameters are determined to be~\cite{Capozzi:2017ipn}
\begin{eqnarray}
\label{eq:3sigma_data}&&\hskip-0.1in 0.250\leq\sin^2\theta_{12}\leq0.354,\quad 0.0190\leq\sin^2\theta_{13}\leq0.0240,\quad
0.381\leq\sin^2\theta_{23}\leq0.615,\\
\nonumber&&\hskip-0,1in6.93\times10^{-5}\text{eV}^2\leq\delta m^2\leq7.96\times10^{-5}\text{eV}^2,\quad
2.411\times10^{-3}\text{eV}^2\leq\Delta m^2\leq2.646\times10^{-3}\text{eV}^2\,,
\end{eqnarray}
for normal ordering (NO) neutrino mass spectrum, and similar results are obtained for inverted ordering (IO) spectrum. Non-Abelian discrete finite groups have been widely used to explain the lepton mixing angles as well as CP violating phases, see Refs.~\cite{Altarelli:2010gt,Ishimori:2010au,King:2013eh,King:2014nza,King:2015aea,King:2017guk} for reviews.

The most appealing possibility for the origin of neutrino mass
seems to be the seesaw mechanism which, in its original formulation, involves heavy right-handed Majorana neutrinos~\cite{seesaw}.
The most minimal version of the seesaw mechanism involves one \cite{King:1998jw} or two right-handed neutrinos~\cite{King:1999mb}. In order to reduce the number of free parameters still further to the smallest number possible, and hence increase predictivity, various
approaches to the two right-handed neutrino seesaw model have been suggested, such as postulating one~\cite{King:2002nf} or two~\cite{Frampton:2002qc} texture zeroes, however such two texture zero models are now phenomenologically excluded~\cite{Harigaya:2012bw} for the case of a normal neutrino mass hierarchy considered here.

The minimal successful seesaw scheme with normal hierarchy is called the Littlest Seesaw (LS) model~\cite{King:2013iva,King:2015dvf,King:2016yvg}, although in fact,
it represents a class of models. The LS models may be defined as two right-handed neutrino models with particularly simple patterns of Dirac mass matrix elements in the basis where both the charged lepton mass matrix and the two-right-handed neutrino mass matrix are diagonal. The Dirac mass matrix typically involves only one texture zero, but the number of parameters is reduced dramatically since each column of this matrix is controlled by a single parameter. In practice this is achieved by introducing a Non-Abelian discrete family symmetry, which is spontaneously broken by flavon fields with particular vacuum alignments governed by remnant subgroups of the family symmetry. Unlike the direct symmetry approach, where a common
residual flavour and remnant CP symmetry is assumed in the
neutrino sector, the Littlest Seesaw approach assumes a {\it different} residual flavour symmetry is preserved by each flavon,
in the diagonal mass basis of two right-handed neutrinos, leading to a highly predictive set of possible alignments.

For example, in the original LS model~\cite{King:2013iva,King:2015dvf,King:2016yvg}, the lepton mixing matrix is predicted to be of the TM1 form in which the first column of the tri-bimaximal mixing matrix is preserved, but with the reactor angle and CP phases fixed by the same two parameters which fix the neutrino masses. This leads to a highly constrained model which is remarkably consistent with current data, but which can be tested in forthcoming neutrino experiments~\cite{Ballett:2016yod}. The LS approach may also be incorporated into grand unified models~\cite{Bjorkeroth:2015ora}. The success of the LS approach, raises the question of whether it is confined to TM1 mixing,
or is of more general applicability. The present paper aims to address this question by considering a different mixing scheme within the same approach, namely the golden ratio (GR) mixing pattern~\cite{GRPrediction1,Kajiyama:2007gx}.

In this paper, we shall propose another viable class of LS models, namely the Golden Littlest seesaw (GLS). Although the golden ratio mixing~\cite{GRPrediction1,Kajiyama:2007gx} is excluded by the measurement of largish reactor mixing angle, the first column of $U_{GR}$ may still be compatible with the experimental data. Inspired by the success of the LS approach for TM1 mixing, we would like to also preserve the first column vector of the GR mixing pattern in our GLS model. We shall perform an exhaustive analysis of all possible remnant symmetries of the group $A_5$ which can be used to enforce various vacuum alignments for the flavon controlling solar mixing,
for two simple cases of the atmospheric flavon vacuum alignment, analogous to the proceedure suggested in the LS approach based on $S_4$. For each possibility we examine the phenomenological viability of the alignment, using figures and extensive benchmark points, comparing our predictions to a recent global analysis of neutrino data. We also repeat the analysis for an alternative form of
Golden Ratio mixing matrix.

The layout of this paper is as follows. In section~\ref{sec:GR} we review GR mixing and the direct model building approach based on the group $A_5$. In section~\ref{sec:LS_GR} we then turn to the GLS approach, based on two right-handed neutrinos with the Dirac mass matrix controlled by flavon vacuum alignments which respect various
remnant symmetries of $A_5$, and examine the phenomenological viability of each case for a discrete choice of phase parameters. In section~\ref{sec:possible_LS} we repeat the procedure for an alternative choice of GR matrix. Section~\ref{sec:introduction} concludes the paper. We report the group theory of $A_5$ in Appendix~\ref{sec:appendix_A5_group}, and the technique details of diagonalizing a two dimensional symmetric matrix are shown in Appendix~\ref{sec:appendix_B}.

\section{\label{sec:GR}Golden Ratio Mixing }

\subsection{Mixing matrix and Klein symmetry}

Before the measurement of the reactor mixing angle, the golden ratio (GR) mixing pattern~\cite{GRPrediction1,Kajiyama:2007gx} was a good leading order approximation and it predicted a zero reactor angle $\theta_{13}=0$, maximal atmospheric mixing angle $\theta_{23}=45^{\circ}$ and a solar mixing angle given by $\cot\theta_{12}=\phi$, where $\phi=(1+\sqrt{5})/2$ is the golden ratio. Note that the golden ratio mixing differs from the tri-bimaximal mixing in the prediction for the solar mixing angle. The explicit form of the golden mixing matrix is given by
\begin{equation}
\label{eq:UGR}
U_{GR}=\begin{pmatrix}
-\sqrt{\frac{\phi}{\sqrt{5}}}  & ~\sqrt{\frac{1}{\sqrt{5}\phi}} ~ & 0  \\
\sqrt{\frac{1}{2\sqrt{5}\phi}}   & ~ \sqrt{\frac{\phi}{2\sqrt{5}}} ~  & -\frac{1}{\sqrt{2}}  \\
\sqrt{\frac{1}{2\sqrt{5}\phi}}   & ~ \sqrt{\frac{\phi}{2\sqrt{5}}} ~  & \frac{1}{\sqrt{2}}
\end{pmatrix}\,.
\end{equation}
In the flavor basis where the charged lepton mass matrix $m_{l}$ is diagonal with $m_{l}=\text{diag}(m_e, m_{\mu}, m_{\tau})$, then the most general form of the neutrino matrix $m_{\nu}$ for the golden ratio mixing is
\begin{equation}
m_{\nu}=U_{GR}\text{diag}(m_1, m_2, m_3)U_{GR}^T= m_1\Phi_1 \Phi_1^T + m_2\Phi_2 \Phi_2^T + m_3\Phi_3 \Phi_3^T\,,
\end{equation}
where the light neutrino masses $m_{1,2,3}$ absorbing the Majorana phases are generally complex, and the vectors $\Phi_{1,2,3}$ are defined as
\begin{equation}
\label{eq:Phi_123}
\Phi_1=\sqrt{\frac{1}{2\sqrt{5}\phi}}\begin{pmatrix}
-\sqrt{2}\phi \\
1\\
1
\end{pmatrix},\qquad \Phi_2=\sqrt{\frac{1}{2\sqrt{5}\phi}}\begin{pmatrix}
\sqrt{2} \\
\phi \\
\phi
\end{pmatrix},\qquad \Phi_3=\frac{1}{\sqrt{2}}\begin{pmatrix}
0  \\
-1  \\
1
\end{pmatrix}\,.
\end{equation}
A unitary transformation $\nu_{L}\rightarrow G_{\nu}\nu_{L}$ of the left-handed Majorana neutrino fields leads to the transformation of the neutrino mass matrix $m_{\nu}\rightarrow G^{T}_{\nu}m_{\nu}G_{\nu}$.
We can check that the above golden ratio neutrino mass matrix is invariant under the following transformations
\begin{equation}
G^{T}_{\nu_i}m_{\nu}G_{\nu_i}=m_{\nu},\quad i=1,2,3\,,
\end{equation}
with
\begin{eqnarray}
G_{\nu_1}=2\Phi_1\Phi^{\dagger}_1-1,\quad G_{\nu_2}=2\Phi_2\Phi^{\dagger}_2-1,\quad G_{\nu_3}=2\Phi_3\Phi^{\dagger}_3-1
\end{eqnarray}
The three flavor symmetry transformations $G_{\nu_1}$, $G_{\nu_2}$ and $G_{\nu_3}$ form a Klein group $K_4\cong Z_2\times Z_2$ and they fulfill
\begin{equation}
\label{k4} G^2_{\nu_i}=\mathbf{1},~~~~~~~G_{\nu_i}G_{\nu_j}=G_{\nu_j}G_{\nu_i}=G_{\nu_k}~~\text{with}~~ i\neq j\neq k\,.
\end{equation}
Furthermore, the symmetry transformation $G_{l}$ of the charged lepton mass matrix is determined by the condition $G^{\dagger}_{l}m^{\dagger}_{l}{m}_{l}G_{l}=m^{\dagger}_{l}{m}_{l}$, therefore $G_{l}$ has to be a diagonal phase matrix. If we choose $G_{l}=\text{diag}(1, \rho, \rho^4)$ with $\rho=e^{2\pi i/5}$, the matrices $G_{l}$, $G_{\nu_1}$, $G_{\nu_2}$ and $G_{\nu_3}$ would give rise to the group $A_5$ in the triplet representation~\cite{Ding:2011cm}.
According to the direct model building approach~\cite{King:2013eh}, if the flavor symmetry $A_5$ is broken to a $Z_5$ subgroup in the charged lepton sector and to Klein subgroup in the neutrino sector, the golden ratio mixing pattern would be obtained naturally~\cite{Ding:2011cm}.

\subsection{\label{subsec:direct_GR} Direct approach in $A_5$}

In both the direct approach and indirect approach, the basis principle of the flavor symmetry model building is the same, that is the different sectors of the Lagrangian preserve different residual subgroups of the flavor symmetry while the whole Lagrangian completely breaks the flavor symmetry. In order to more clearly understand the idea of the GLS, we shall briefly recapitulate the direct approach to the GR mixing from $A_5$ flavor symmetry before presenting our GLS
within the indirect approach in the following section.

\begin{table}[hptb!]
\begin{center}
\begin{tabular}{|c|c|c|}\hline\hline
 ~~  &  $S$  &   $T$     \\ \hline
~~~${\bf 1}$ ~~~ & 1   &  1    \\ \hline
   &   &      \\ [-0.16in]
${\bf 3}$ &  $\frac{1}{\sqrt{5}}
\begin{pmatrix}
 1 &~ -\sqrt{2} &~ -\sqrt{2} \\
 -\sqrt{2} &~ -\phi  &~ 1/\phi \\
 -\sqrt{2} &~ 1/\phi &~ -\phi
\end{pmatrix} $
    & $\begin{pmatrix}
 1 &~ 0 &~ 0 \\
 0 &~ \rho  &~ 0 \\
 0 &~ 0 &~ \rho^4
\end{pmatrix} $
   \\ [0.12in]\hline
   &   &     \\ [-0.16in]

${\bf 3}^{\prime}$ &  $~\frac{1}{\sqrt{5}}
\begin{pmatrix}
 -1 &~ \sqrt{2} &~ \sqrt{2} \\
 \sqrt{2} &~ -1/\phi &~ \phi  \\
 \sqrt{2} &~ \phi  &~ -1/\phi
\end{pmatrix}$
    & $\begin{pmatrix}
 1 &~ 0 &~ 0 \\
 0 &~ \rho^2 &~ 0 \\
 0 &~ 0 &~ \rho^3
\end{pmatrix}$
   \\ [0.12in]\hline
   &   &     \\ [-0.16in]

${\bf 4}$  & $\frac{1}{\sqrt{5}}
\begin{pmatrix}
 1 &~ 1/\phi &~ \phi  &~ -1 \\
1/\phi &~ -1 &~ 1 &~ \phi  \\
 \phi  &~ 1 &~ -1 &~ 1/\phi \\
 -1 &~ \phi  &~ 1/\phi &~ 1
\end{pmatrix}$
    & $\begin{pmatrix}
 \rho  &~ 0 &~ 0 &~ 0 \\
 0 &~ \rho^2 &~ 0 &~ 0 \\
 0 &~ 0 &~ \rho^3 &~ 0 \\
 0 &~ 0 &~ 0 &~ \rho^4
\end{pmatrix}$ \\ [0.12in]\hline
   &   &     \\ [-0.16in]

${\bf 5}$  & $\frac{1}{5}
\begin{pmatrix}
 -1 &~ \sqrt{6} &~ \sqrt{6} &~ \sqrt{6} &~ \sqrt{6} \\
 \sqrt{6} &~ 1/\phi^{2} &~ -2 \phi  &~ 2/\phi &~ \phi^2 \\
 \sqrt{6} &~ -2\phi  &~ \phi^2&~ 1/\phi^{2} &~ 2/\phi \\
 \sqrt{6} &~ 2/\phi &~ 1/\phi^{2} &~ \phi^2 &~ -2\phi  \\
 \sqrt{6} &~ \phi^2 &~ 2/\phi &~ -2\phi  &~ 1/\phi^{2}
\end{pmatrix}$
    & $\begin{pmatrix}
 1 &~ 0 &~ 0 &~ 0 &~ 0 \\
 0 &~ \rho  &~ 0 &~ 0 &~ 0 \\
 0 &~ 0 &~ \rho^2 &~ 0 &~ 0 \\
 0 &~ 0 &~ 0 &~ \rho^3 &~ 0 \\
 0 &~ 0 &~ 0 &~ 0 &~ \rho^4
\end{pmatrix}$
\\[0.22in] \hline\hline
\end{tabular}
\caption{\label{tab:representation_Td}The representation matrices of the generators $S$ and $T$ for the five irreducible representations of $A_5$ group in the basis which is convenient for discussing the Golden ratio mixing pattern, where $\rho=e^{2\pi i/5}$ is the fifth root of unit.}
\end{center}
\end{table}

We first recall that $A_5$ is the even permutation group of five objects. Geometrically $A_5$ is the symmetry group of the icosahedron. The $A_5$ group can be generated by two generators $S$ and $T$ which satisfy the following multiplication rules
\begin{equation}
S^2=T^5=(ST)^3=1\,.
\end{equation}
The $A_5$ group has five irreducible representations: one single $\mathbf{1}$, two triplets $\mathbf{3}$ and $\mathbf{3}'$,  one four-dimensional representation $\mathbf{4}$ and one five-dimensional representation $\mathbf{5}$. The explicit form of the representation matrices for the generators $S$ and $T$ are collected in table~\ref{tab:representation_Td}. The interested readers can refer to Ref.~\cite{Li:2015jxa} for detailed group theory of $A_5$ and Clebsch-Gordan coefficients. The interplay between $A_5$ flavor symmetry and lepton mixing has been extensively studied in the literature~\cite{Ding:2011cm,Li:2015jxa,deAdelhartToorop:2011re,Feruglio:2011qq,Everett:2008et,Cooper:2012bd,DiIura:2015kfa,Ballett:2015wia}. We find that the representation matrices of the generators $S$ and $T$ in $\mathbf{3}'$ exactly coincide with those of $T^3ST^2ST^3$ and $T^2$ respectively in $\mathbf{3}$. This implies that the set of all matrices describing the representations $\mathbf{3}$ and $\mathbf{3}'$ are the same. Therefore the same results would be obtained no matter if the left-handed leptons transform as $\mathbf{3}$ or $\mathbf{3}'$ of $A_5$. Without loss of generality, We shall assign the three generations of left-handed leptons to the triplet $\mathbf{3}$ in the following.

In the direct approach, the $A_5$ flavor symmetry group is assumed to be broken to a abelian subgroup $G_{l}$ such as $G_{l}=Z^{T}_5$. As a consequence the charged lepton mass matrix $m_{l}$ is invariant under the action of the element $T$, i.e.
\begin{equation}
\rho^{\dagger}_{\mathbf{3}}(T)m^{\dagger}_{l}m_{l}\rho_{\mathbf{3}}(T)=m^{\dagger}_{l}m_{l}\,.
\end{equation}
This implies that the unitary transformation $U_{l}$ which diagonalizes the charged lepton mass matrix $U^{\dagger}_{l}(T)m^{\dagger}_{l}m_{l}U_{l}=\text{diag}(m^2_{e}, m^2_{\mu}, m^2_{\tau})$ has the property
\begin{equation}
U^{\dagger}_{l}\rho_{\mathbf{3}}(T)U_{l}=\text{diag}(1, e^{i\frac{2\pi}{5}}, -e^{i\frac{3\pi}{5}})\,.
\end{equation}
Since the generator $T$ is diagonal with $\rho_{\mathbf{3}}(T)=\text{diag}(1, e^{i\frac{2\pi}{5}}, -e^{i\frac{3\pi}{5}})$ in our working basis, $U_{l}$ has to be a unit matrix,
\begin{equation}
U_{l}=\begin{pmatrix}
1  ~&~   0  ~&~ 0 \\
0  ~&~  1   ~&~  0 \\
0  ~&~  0   ~&~  1
\end{pmatrix}\,.
\end{equation}
As a consequence, the lepton mixing completely arises from the neutrino mixing. The $A_5$ flavor symmetry is broken down to a Klein subgroup $G_{\nu}$ in the neutrino sector in the paradigm of direct approach. Here we choose $G_{\nu}=K^{(S, T^3ST^2ST^3)}_4$ whose representation matrices in the chosen basis are
\begin{eqnarray}
\nonumber&&\rho_{\mathbf{3}}(S)=\frac{1}{\sqrt{5}}
\begin{pmatrix}
 1 &~ -\sqrt{2} &~ -\sqrt{2} \\
 -\sqrt{2} &~ -\phi  &~ 1/\phi \\
 -\sqrt{2} &~ 1/\phi &~ -\phi
\end{pmatrix},\\
\nonumber&&\rho_{\mathbf{3}}(T^3ST^2ST^3)=\frac{1}{\sqrt{5}}\begin{pmatrix}
-1  & ~  \sqrt{2}  &~  \sqrt{2} \\
 \sqrt{2}  &~ -1/\phi   &~  \phi\\
  \sqrt{2} &~   \phi   &~  -1/\phi
\end{pmatrix}\\
&&\rho_{\mathbf{3}}(T^3ST^2ST^3S)=-\begin{pmatrix}
1  &~ 0  &~ 0  \\
0  &~ 0  &~  1  \\
0  &~ 1  &~ 0
\end{pmatrix}\,,
\end{eqnarray}
Then the neutrino diagonalization matrix $U_{\nu}$ turns out to be the golden ratio mixing pattern~\cite{Ding:2011cm,Li:2015jxa,deAdelhartToorop:2011re,Feruglio:2011qq},
\begin{equation}
U_{GR}=\begin{pmatrix}
-\sqrt{\frac{\phi}{\sqrt{5}}}  & ~\sqrt{\frac{1}{\sqrt{5}\phi}} ~ & 0  \\
\sqrt{\frac{1}{2\sqrt{5}\phi}}   & ~ \sqrt{\frac{\phi}{2\sqrt{5}}} ~  & -\frac{1}{\sqrt{2}}  \\
\sqrt{\frac{1}{2\sqrt{5}\phi}}   & ~ \sqrt{\frac{\phi}{2\sqrt{5}}} ~  & \frac{1}{\sqrt{2}}
\end{pmatrix}\,.
\end{equation}
Notice that the three column vectors $\Phi_{1,2,3}$ of the GR mixing preserve three different $Z_2$ subgroups of $A_5$,
\begin{equation}
\rho_{\mathbf{3}}(S)\Phi_1=\Phi_1,\quad \rho_{\mathbf{3}}(T^3ST^2ST^3)\Phi_2=\Phi_2,\quad \rho_{\mathbf{3}}(T^3ST^2ST^3S)\Phi_3=\Phi_3\,.
\end{equation}
We summarize that the GR mixing arises from the mismatch between the residual subgroups $G_{l}$ and $G_{\nu}$ in the direct approach.

\section{\label{sec:LS_GR}Golden Littlest Seesaw in $A_5$}

\subsection{Littlest Seesaw}
\begin{figure}[t!]
\centering
\includegraphics[width=0.55\textwidth]{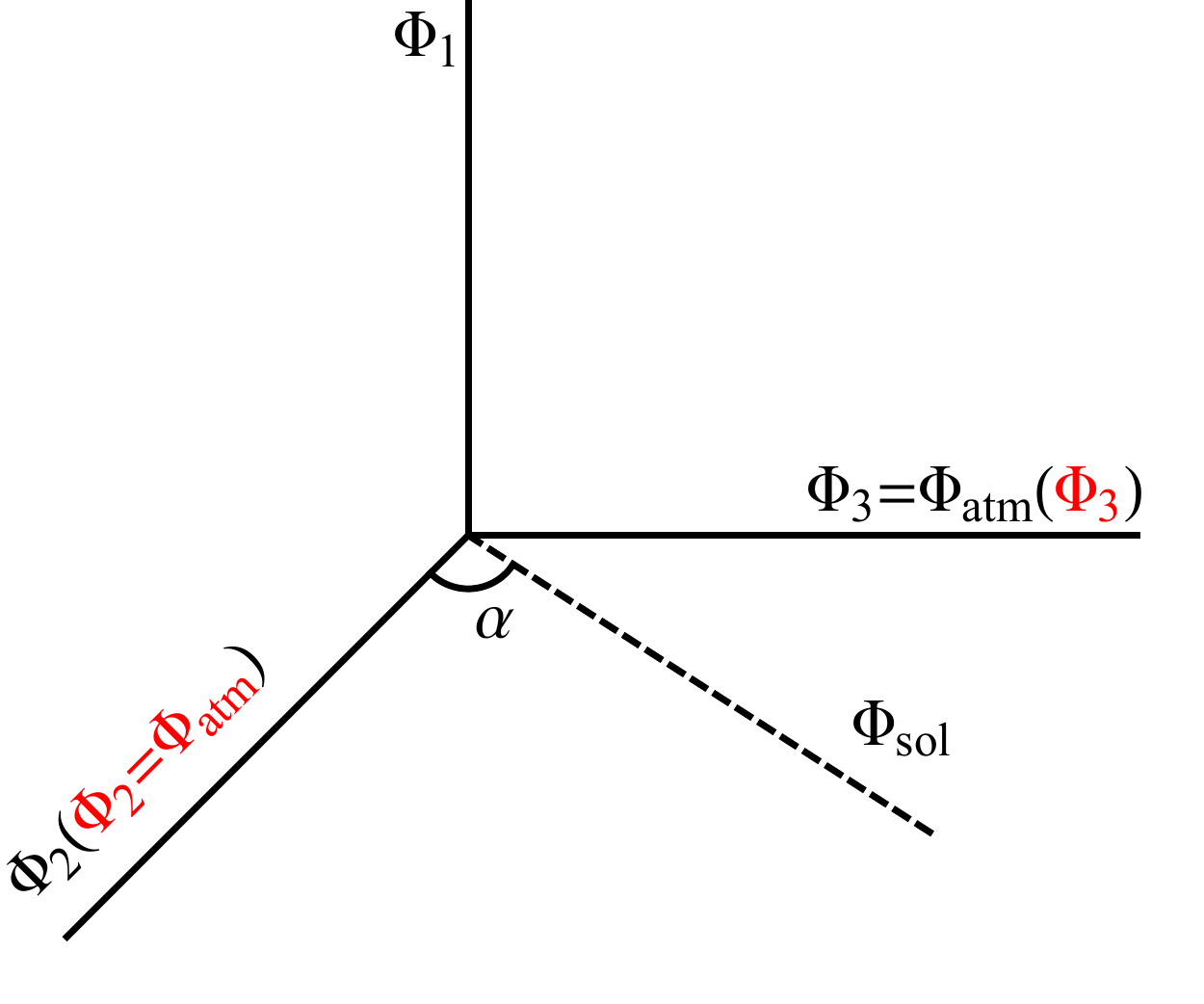}
\caption{\label{fig:alignment} The vacuum alignment in the Littlest seesaw model. $\Phi_{1}$, $\Phi_{2}$ and $\Phi_{3}$ are the three columns of the golden ratio mixing matrix. The alignment vector $\Phi_{\text{atm}}$ is either $\Phi_{2}$ or $\Phi_{3}$, and $\Phi_{\text{sol}}$ is a general vector orthogonal to $\Phi_{1}$.}
\end{figure}

The indirect model building approach~\cite{King:2013eh}
is an interesting alternative to the direct approach. In the indirect approach, the original flavor symmetry is completely broken in the neutrino sector, and the residual symmetry $Z_2\times Z_2$ of the neutrino mass matrix arises accidentally. The basic idea of the indirect approach is to effectively promote the columns of the Dirac mass matrix to fields which transform as triplets under the flavour symmetry. We assume that the Dirac mass matrix can be written as $m_D= (a\Phi_{\rm atm},b\Phi_{\rm sol},c\Phi_{\rm dec})$ where the columns are proportional to triplet Higgs scalar fields with
particular vacuum alignments and $a,b,c$ are three constants of proportionality. It is convenient to work in the basis where the right-handed neutrino mass matrix are diagonal with the mass eigenvalues equal to $M_{\rm atm}$, $M_{\rm sol}$ and $M_{\rm dec}$.
Then the light neutrino mass matrix given by the seesaw formula is
\begin{equation}
\label{eq:mnu_indirect}m_{\nu}=a^2\frac{\Phi_{\rm atm}\Phi_{\rm atm}^T}{M_{\rm atm}}
+ b^2\frac{\Phi_{\rm sol}\Phi_{\rm sol}^T}{M_{\rm sol}}
+ c^2\frac{\Phi_{\rm dec}\Phi_{\rm dec}^T}{M_{\rm dec}}\,,
\end{equation}
where we have dropped an overall minus sign which is physically irrelevant. In the case that the columns of the Dirac mass matrix are proportional to the columns of the GR matrix, $\Phi_{\rm atm}\propto \Phi_3$, $\Phi_{\rm sol}\propto \Phi_2$ and $\Phi_{\rm dec}\propto \Phi_1$, the three columns of $m_{D}$ would be mutually orthogonal, as illustrated in figure~\ref{fig:alignment}.
As a consequence, the resulting effective light Majorana mass matrix
$m_{\nu}$ is form diagonalizable, and it is exactly diagonalized by the golden ratio mixing matrix. This scenario is referred to as form dominance~\cite{Chen:2009um}. In the limit of $M_{\rm dec}\gg M_{\rm atm}, M_{\rm sol}$, as a good leading order approximation we could drop the last term and the model reduces to a two right-handed neutrino model, such that the lightest neutrino is massless.

The Littlest seesaw framework assumes that there are only two right-handed neutrinos to begin with, together with flavons which couple to them with particular vacuum alignments, leading to
the columns of the Dirac mass matrix taking the above forms.
Within the Littlest seesaw framework~\cite{King:2015dvf}, we shall assume that both vacuum alignments $\Phi_{\text{sol}}$ and $\Phi_{\text{atm}}$ are orthogonal to $\Phi_1$, in order to preserve the first column of the mixing matrix. Then we shall choose $\Phi_{\text{atm}}$ to be either $\Phi_2$ or $\Phi_3$, and take $\Phi_{\text{sol}}$ to be a general vector orthogonal to $\Phi_1$, as illustrated in figure~\ref{fig:alignment}. Later on we shall fix the alignment of $\Phi_{\text{sol}}$ by appealing to remnant symmetry,
according to a generalisation of the direct approach, as discussed in the next subsection.

The Littlest seesaw is clearly a rather predictive framework which combines the two right-handed neutrino model with the indirect approach~\cite{King:1998jw}. In this framework, two right-handed neutrinos $N^{\mathrm{atm}}_R$ and $N^{\mathrm{sol}}_R$ are introduced, and the third right-handed neutrino is assumed to be almost decoupled and irrelevant. $N^{\mathrm{atm}}_R$ dominantly contributes to the seesaw mechanism and is mainly responsible for the atmospheric neutrino mass $m_3$. $N^{\mathrm{sol}}_R$ is sub-dominant and is mainly responsible for the solar neutrino mass $m_2$ while the lightest neutrino mass $m_1$ is zero in this limit. The Littlest seesaw model generally assumes three generations of left-handed neutrino fields $\nu_{L}=\left(\nu_{e}, \nu_{\mu}, \nu_{\tau}\right)$ transforms as a triplet of the flavor symmetry while both $N^{\mathrm{atm}}_R$ and $N^{\mathrm{sol}}_R$ are singlets. In the flavor basis where the charged lepton mass matrix is diagonal with real positive eigenvalues $m_{e}$, $m_{\mu}$, $m_{\tau}$ and the right-handed neutrino Majorana mass matrix is also diagonal, by introducing appropriate auxiliary abelian symmetry,  the
generic Littlest seesaw Lagrangian can be written as
\begin{small}
\begin{equation}
\mathcal{L}=-y_{\mathrm{atm}}\bar{L}.\phi_{\mathrm{atm}}N_R^{\mathrm{atm}}-y_{\mathrm{sol}}\bar{L}.\phi_{\mathrm{sol}}N_R^{\mathrm{sol}}
-\frac{1}{2}M_{\mathrm{atm}}\overline{(N^{\mathrm{atm}}_R)^c}N_R^{\mathrm{atm}}-\frac{1}{2}M_{\mathrm{sol}}\overline{(N^{\mathrm{sol}}_R)^c}N_R^{\mathrm{sol}}
+{h.c.}\,,
\end{equation}
\end{small}
where the flavons $\phi_{\rm sol}$ and $\phi_{\rm atm}$ can be either Higgs fields transforming as triplets under the flavour symmetry, or combinations of a single Higgs electroweak doublet together with triplet flavons. The fields $L$ are the electroweak lepton doublets which are unified into a triplet representation of the flavor symmetry group. Then $\Phi_{\mathrm{atm}}$ and $\Phi_{\mathrm{sol}}$ in Eq.~\eqref{eq:mnu_indirect} arise from the vacuum expectation values (VEVs) of $\phi_{\rm sol}$ and $\phi_{\rm atm}$ respectively.

\subsection{\label{subsec:indirect_GR1}Indirect approach in $A_5$}

\begin{figure}[t!]
\centering
\begin{tabular}{c}
\includegraphics[width=0.50\linewidth]{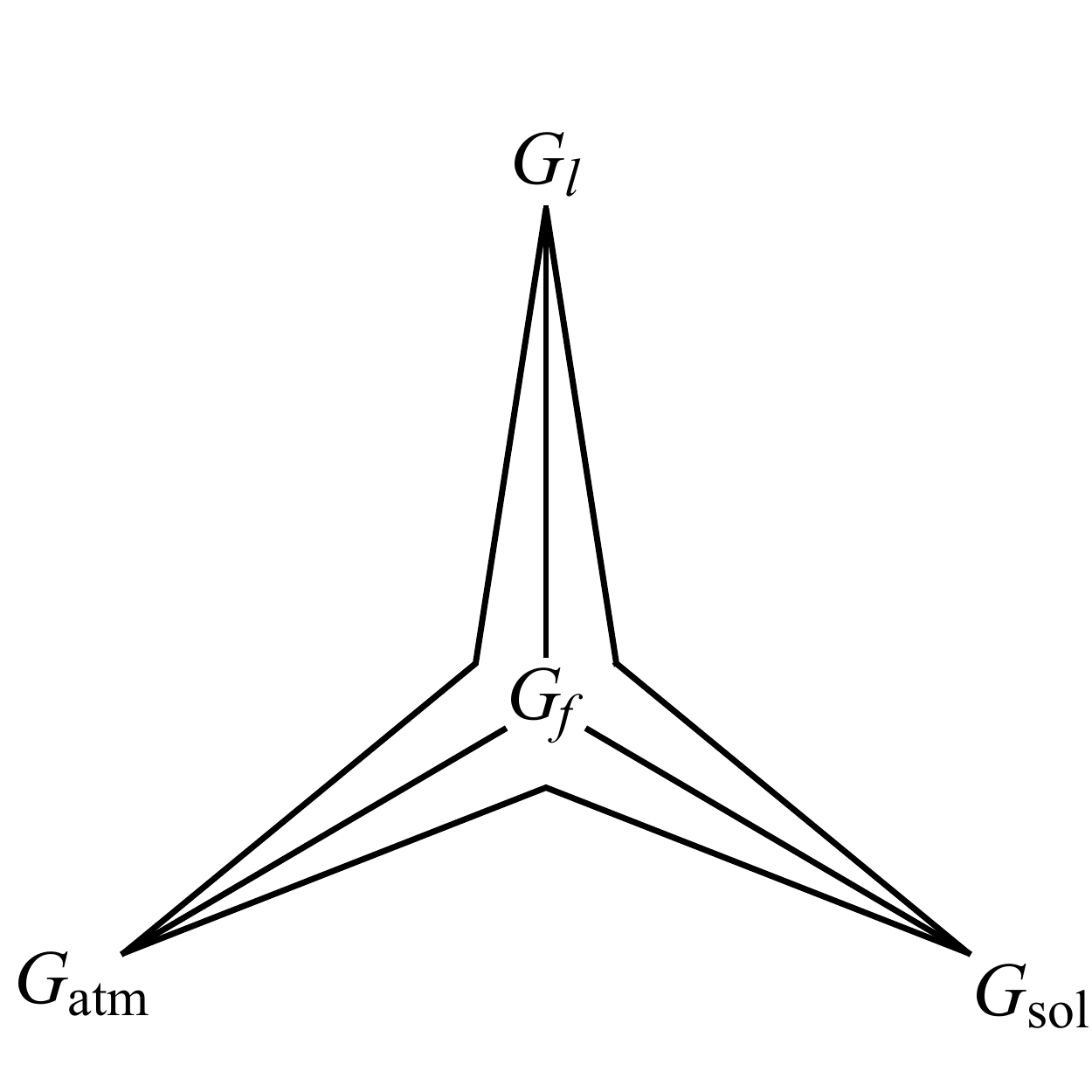}
\end{tabular}
\caption{\label{fig:benz} A sketch of the indirect model building approach, where the charged lepton preserves a residual subgroup $G_{l}$, and the neutrino vacuum alignments $\Phi_{\text{atm}}$ and $\Phi_{\text{sol}}$ are enforced by the residual symmetries $G_{\text{atm}}$ and $G_{\text{sol}}$ respectively.
}
\end{figure}

The indirect approach is a further generalization of the direct approach. We assume that the $A_5$ group is broken to the abelian subgroup  $G_{l}=Z^{T}_5$ in the charged lepton sector, the vacuum alignments $\Phi_{\text{atm}}$ and $\Phi_{\text{sol}}$ preserve different residual symmetries $G_{\text{atm}}$ and $G_{\text{sol}}$ respectively while the $A_5$ flavor symmetry is completely broken in the entire neutrino sector. The indirect approach is schematically illustrated in figure~\ref{fig:benz}. In our GLS model, as stated above the alignment vector $\Phi_{\text{sol}}$ is orthogonal to $\Phi_1$, its most general form is
\begin{equation}
\label{eq:phi_sol}
\Phi_{\text{sol}}\propto\left(\sqrt{2},~ \phi-x,~\phi+x\right)^{T}\,.
\end{equation}
We find there are five possible values of $x$ related to certain residual subgroups of $A_5$,
\begin{equation}
\label{eq:phi_sol_res_sym}
\begin{aligned}
x&=0,\qquad G_{\text{sol}}=Z^{T^3ST^2ST^3}_2,  \\
x&=2i\phi^2\sin\frac{2\pi}{5},\qquad G_{\text{sol}}=Z^{T^3ST^2S}_3,\\
x&=-2i\phi^2\sin\frac{2\pi}{5},\qquad G_{\text{sol}}=Z^{ST^2ST^3}_3,\\
x&=-2i\sin\frac{\pi}{5},\qquad G_{\text{sol}}=Z^{T^2ST}_5,\\
x&=2i\sin\frac{\pi}{5},\qquad G_{\text{sol}}=Z^{TST^2}_5\,.
\end{aligned}
\end{equation}
Accordingly the vacuum alignment of the solar flavon $\phi_{\text{sol}}$ is:
\begin{equation}
\begin{aligned}
x&=0,\qquad \Phi_{\text{sol}}=\left(\sqrt{2},~ \phi,~\phi\right)^{T},  \\
x&=2i\phi^2\sin\frac{2\pi}{5},\qquad \Phi_{\text{sol}}=\left(\sqrt{2},~ 2\phi^2e^{-2i\pi/5},~2\phi^2e^{2i\pi/5}\right)^{T},\\
x&=-2i\phi^2\sin\frac{2\pi}{5},\qquad \Phi_{\text{sol}}=\left(\sqrt{2},~ 2\phi^2e^{2i\pi/5},~2\phi^2e^{-2i\pi/5}\right)^{T},\\
x&=-2i\sin\frac{\pi}{5},\qquad \Phi_{\text{sol}}=\left(\sqrt{2},~ 2e^{i\pi/5},~2e^{-i\pi/5}\right)^{T},\\
x&=2i\sin\frac{\pi}{5},\qquad \Phi_{\text{sol}}=\left(\sqrt{2},~ 2e^{-i\pi/5},~2e^{i\pi/5}\right)^{T}\,.
\end{aligned}
\end{equation}
In our framework, another alignment vector $\Phi_{\text{atm}}$ is assumed to be along the direction of $\Phi_3$ or $\Phi_2$. In the following, we shall discuss the two cases one by one.

\subsubsection{\label{subsubsec:UGR1_3rd_col_Td}Golden Littlest seesaw with $\Phi_{\mathrm{atm}}\propto\Phi_{3}$ }

\begin{figure}[hptb]
\centering
\begin{tabular}{c}
\includegraphics[width=0.99\linewidth]{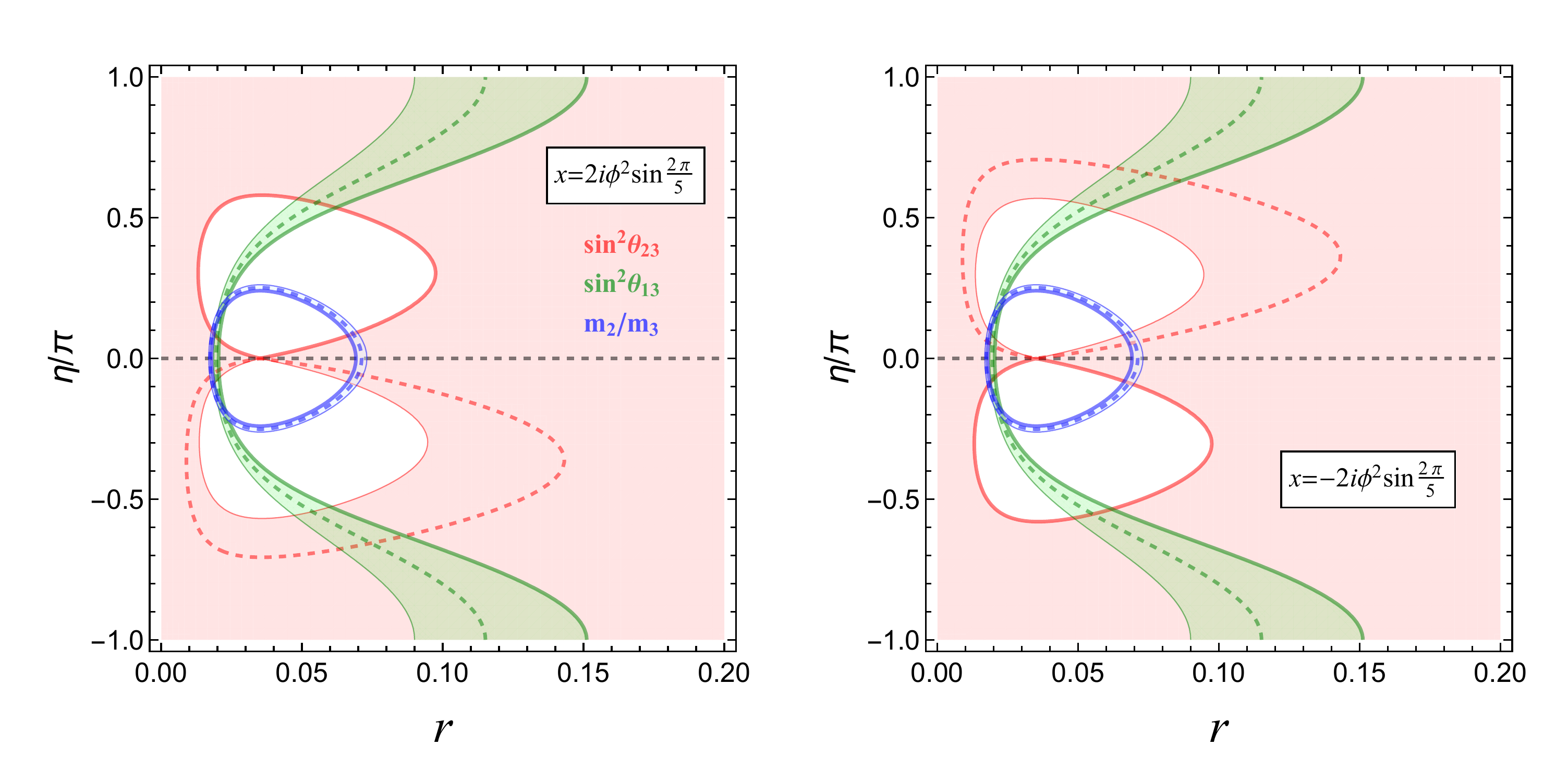}
\end{tabular}
\caption{\label{UGR1_3rd_col_Td} Contour plots of $\sin^2\theta_{13}$, $\sin^2\theta_{23}$ and $m_2/m_3$ in the $\eta-r$ plane for the golden Littlest seesaw with $\Phi_{\mathrm{atm}}\propto\Phi_{3}$. Here we take $x=2i\phi^2\sin(2\pi/5)$ and $x=-2i\phi^2\sin(2\pi/5)$ for which the solar vacuum alignment $\Phi_{\text{sol}}$ preserves the residual symmetry $G_{\text{sol}}=Z^{T^3ST^2S}_3$ and $G_{\text{sol}}=Z^{ST^2ST^3}_3$ respectively. The $3\sigma$ upper (lower) bounds of the lepton mixing angles are labelled with thick (thin) solid curves, and the dashed contour lines represent the corresponding best fit values. The $3\sigma$ ranges as well as the best fit values of the mixing angles are adapted from~\cite{Capozzi:2017ipn}. The black contour line refers to maximal atmospheric mixing angle with $\sin^2\theta_{23}=0.5$.
}
\end{figure}

In this case, the vacuum $\Phi_{\text{atm}}$ reads as
\begin{equation}
\Phi_{\text{atm}}\propto \left(0,~-1,~1\right)^{T}\,,
\end{equation}
which is invariant under the action of the $Z^{T^3ST^2ST^3S}_2$ subgroup. Consequently the Dirac neutrino mass matrix $M_D$ and the right-handed neutrino heavy Majorana mass matrix $M_{N}$ are given by
\begin{equation}
M_{D}=\begin{pmatrix}
0 ~&~   \sqrt{2} b \\
-a   ~&~  (\phi-x) b \\
a  ~&~  (\phi +x)b
\end{pmatrix}\,,
\qquad M_{N}=\begin{pmatrix}
M_{\mathrm{atm}}  &  0  \\
0  &   M_{\mathrm{sol}}
\end{pmatrix}
\end{equation}
Integrating out the right-handed neutrinos, the light effective Majorana neutrino mass matrix is approximately given by the seesaw formula
\begin{eqnarray}
\nonumber m_{\nu}&=&-M_{D}M^{-1}_{N}M^{T}_{D}\\
\label{eq:mnu_GR_3col_Td} &=&m_a\begin{pmatrix}
 0 &~ 0 ~& 0 \\
 0 &~ 1 ~& -1 \\
 0 &~ -1 ~& 1 \\
\end{pmatrix}+m_{b}e^{i\eta}
\begin{pmatrix}
 2 &~ \sqrt{2} (\phi -x) ~& \sqrt{2} (x+\phi ) \\
 \sqrt{2} (\phi -x) &~ (x-\phi )^2 ~& -x^2+\phi +1 \\
 \sqrt{2} (x+\phi ) &~ -x^2+\phi +1 ~& (x+\phi )^2 \\
\end{pmatrix}\,,
\end{eqnarray}
where $m_a=|a|^2/M_{\mathrm{atm}}$, $m_b=|b|^2/M_{\mathrm{sol}}$, the relative phase $\eta=\mathrm{arg}(b^2/a^2)$, and an overall phase of $m_{\nu}$ has been omitted. Therefore four parameters $m_a$, $m_{b}$, $x$ and $\eta$ describe both the neutrino flavor mixing and neutrino masses. One can check that neutrino mass matrix $m_{\nu}$ of Eq.~\eqref{eq:mnu_GR_3col_Td} satisfies
\begin{equation}
m_{\nu}\left(\begin{array}{c}
-\sqrt{\frac{\phi}{\sqrt{5}}} \\
\sqrt{\frac{1}{2\sqrt{5}\phi}}\\
\sqrt{\frac{1}{2\sqrt{5}\phi}}
\end{array}
\right)=\left(\begin{array}{c}
0\\
0\\
0
\end{array}\right)\,.
\end{equation}
This implies that the column vector $(-\sqrt{\frac{\phi}{\sqrt{5}}}, \sqrt{\frac{1}{2\sqrt{5}\phi}}, \sqrt{\frac{1}{2\sqrt{5}\phi}})^T$ is an eigenvector of $m_\nu$ with a zero eigenvalue. As a result, the first column of the PMNS mixing matrix exactly coincides with the GR mixing pattern, and the corresponding light neutrino mass vanishes $m_1=0$. In order to diagonalize the above neutrino mass matrix, we firstly perform a golden ratio transformation and obtain
\begin{equation}
\label{eq:mnu_UGR1_Td}
m^{\prime}_{\nu}=U^{T}_{GR}m_{\nu}U_{GR}=\begin{pmatrix}
0  &~   0  &~  0 \\
0  &~  y   &~ z  \\
0  &~  z  &~ w
\end{pmatrix}
\end{equation}
where
\begin{eqnarray}
\nonumber&&y=2\sqrt{5}\,\phi m_{b}\,e^{i\eta},\\
 \nonumber&&z=2x\sqrt{\phi+2}\,m_{b}\,e^{i\eta},\\
\label{eq:nu_mass_par_UGR1_Td}  &&w=|w|e^{i\phi_{w}}=2(m_a+x^2\,m_{b}\,e^{i\eta})\,.
\end{eqnarray}
The neutrino mass matrix $m_{\nu}$ in Eq.~\eqref{eq:mnu_UGR1_Td} by diagonalized through the standard procedure, as shown in Ref.~\cite{Ding:2013bpa}. We have
\begin{equation}
{U^{\prime}}^{T}_{\nu}m^{\prime}_{\nu}U^{\prime}_{\nu}=\text{diag}(0, m_2, m_3)\,,
\end{equation}
where the unitary matrix $U^{\prime}_{\nu}$ can be written as
\begin{equation}
\label{eq:Uprime}U^{\prime}_{\nu}=\begin{pmatrix}
1  &~    0    &~   0 \\
0  &~  \cos\theta \,e^{i(\psi+\rho)/2}   &~  \sin\theta \,e^{i(\psi+\sigma)/2}   \\
0  &~   -\sin\theta\, e^{i(-\psi+\rho)/2}    &~~    \cos\theta\, e^{i(-\psi+\sigma)/2}
\end{pmatrix}\,.
\end{equation}
We find the light neutrino masses $m_{2,3}$ are
\begin{eqnarray}
\nonumber&&m^2_2=\frac{1}{2}\left[|y|^2+|w|^2+2|z|^2-\frac{|w|^2-|y|^2}{\cos2\theta}\right]\,,\\
\label{eq:nu_mass_UGR1}&&m^2_3=\frac{1}{2}\left[|y|^2+|w|^2+2|z|^2+\frac{|w|^2-|y|^2}{\cos2\theta}\right]
\end{eqnarray}
The rotation angle $\theta$ is determined to be
\begin{eqnarray}
\nonumber&&\sin2\theta=\frac{-2iz\,e^{-i\eta}\sqrt{|y|^2+|w|^2-2|y||w|\cos(\phi_{w}-\eta)}}{\sqrt{(|w|^2-|y|^2)^2+4|z|^2\left[|y|^2+|w|^2-2|y||w|\cos(\phi_{w}-\eta)\right]}},\\
\label{eq:theta_UGR1_Td}&&\cos2\theta=\frac{|w|^2-|y|^2}{\sqrt{(|w|^2-|y|^2)^2+4|z|^2\left[|y|^2+|w|^2-2|y||w|\cos(\phi_{w}-\eta)\right]}}\,.
\end{eqnarray}
The phases $\psi$, $\rho$ and $\sigma$ are given by
\begin{small}
\begin{eqnarray}
\nonumber&&\hskip-0.08in\sin\psi=\frac{|y|-|w|\cos(\phi_{w}-\eta)}{\sqrt{|y|^2+|w|^2-2|y||w|\cos(\phi_{w}-\eta)}},~ \cos\psi=\frac{|w|\sin(\phi_{w}-\eta)}{\sqrt{|y|^2+|w|^2-2|y||w|\cos(\phi_{w}-\eta)}},\\
\nonumber&&\hskip-0.08in\sin\rho=-\frac{(m^2_2-|z|^2)\cos\eta-|y||w|\cos\phi_{w}}{m_2\sqrt{|y|^2+|w|^2-2|y||w|\cos(\phi_{w}-\eta)}},~
\cos\rho=\frac{-(m^2_2-|z|^2)\sin\eta+|y||w|\sin\phi_{w}}{m_2\sqrt{|y|^2+|w|^2-2|y||w|\cos(\phi_{w}-\eta)}},\\
\label{eq:prs_UGR1_Td}&&\hskip-0.08in\sin\sigma=-\frac{(m^2_3-|z|^2)\cos\eta-|y||w|\cos\phi_w}{m_3\sqrt{|y|^2+|w|^2-2|y||w|\cos(\phi_{w}-\eta)}},~
\cos\sigma=\frac{-(m^2_3-|z|^2)\sin\eta+|y||w|\sin\phi_w}{m_3\sqrt{|y|^2+|w|^2-2|y||w|\cos(\phi_{w}-\eta)}}\,.
\end{eqnarray}
\end{small}
Thus the lepton mixing matrix is determined to be
\begin{equation}\label{eq:PMNS_UGR_3col_Td}
\hskip-0.1in U=U_{GR}U^{\prime}_{\nu}=\sqrt{\frac{1}{2\sqrt{5}\phi}}
\begin{pmatrix}
 -\sqrt{2} \phi  &~ \sqrt{2} \cos \theta  ~& \sqrt{2} e^{i \psi } \sin \theta  \\
 1 &~ \phi  \cos \theta +\sqrt{\phi+2}  \sin \theta\, e^{-i \psi }  ~&  \phi  \sin \theta \,e^{i \psi } -\sqrt{\phi+2}  \cos \theta  \\
 1 &~ \phi  \cos \theta -\sqrt{\phi+2} \sin \theta \, e^{-i \psi } ~&  \phi  \sin \theta \,e^{i \psi } +\sqrt{\phi+2} \cos \theta   \\
\end{pmatrix}P_{\nu}\,,
\end{equation}
with
\begin{equation}
P_{\nu}=\text{diag}(1, e^{i(\psi+\rho)/2}, e^{i(-\psi+\sigma)/2})\,.
\end{equation}
The most general leptonic mixing matrix in the two right-handed neutrino model can be parameterized as
\begin{equation}\label{eq:PMNS_UGR1_Td}
U=\left(\begin{array}{ccc}
c_{12}c_{13}  &   s_{12}c_{13}   &   s_{13}e^{-i\delta_{CP}}  \\
-s_{12}c_{23}-c_{12}s_{13}s_{23}e^{i\delta_{CP}}   &  c_{12}c_{23}-s_{12}s_{13}s_{23}e^{i\delta_{CP}}  &  c_{13}s_{23}  \\
s_{12}s_{23}-c_{12}s_{13}c_{23}e^{i\delta_{CP}}   & -c_{12}s_{23}-s_{12}s_{13}c_{23}e^{i\delta_{CP}}  &  c_{13}c_{23}
\end{array}\right)\text{diag}(1,e^{i\frac{\beta}{2}},1)\,,
\end{equation}
where $c_{ij}\equiv \cos\theta_{ij}$, $s_{ij}\equiv \sin\theta_{ij}$, $\delta_{CP}$ is the Dirac CP violation phase and $\beta$ is the Majorana CP phase. Note that a second Majorana phase is needed if the lightest neutrino is not massless. Then we can extract the expressions for the lepton mixing angles as follows
\begin{eqnarray}
\nonumber &&\sin^2\theta_{13}=\frac{\sin^2\theta}{\sqrt{5}\phi}\,, \quad \sin^2\theta_{12}=\frac{\cos^2\theta }{\sqrt{5}\phi-\sin^2\theta }\,, \\
\label{eq:angles_GR_3col_Td}&& \sin^2\theta_{23}=\frac{1}{2}-\frac{ \sqrt{3+4\phi}\,\sin2\theta\cos\psi}{2(\sqrt{5} \phi-\sin^2\theta)}\,.
\end{eqnarray}
Eliminating the free parameter $\theta$, we see that a sum rule between the solar mixing angle $\theta_{12}$ and the reactor mixing angle $\theta_{13}$ is satisfied,
\begin{equation}
\label{eq:sum_rule_GR3}\cos^2\theta_{12}\cos^2\theta_{13}=\frac{\phi}{\sqrt{5}}\,.
\end{equation}
Using the best fit value of $\sin^2\theta_{13}=0.0215$,  we find for the solar mixing angle
\begin{equation}
\sin^2\theta_{12}\simeq0.261\,,
\end{equation}
which is within the $3\sigma$ region~\cite{Capozzi:2017ipn}. As regards the CP violation, two weak basis invariants $J_{CP}$~\cite{Jarlskog:1985ht} and $I_1$~\cite{Branco:1986gr} associated with the CP phases $\delta_{CP}$ and $\beta$ respectively can be defined,
\begin{eqnarray}
\nonumber && J_{CP}=\Im{(U_{11}U_{33}U^{*}_{13}U^{*}_{31})}=\frac{1}{8}\sin2\theta_{12}\sin2\theta_{13}\sin2\theta_{23}\cos\theta_{13}\sin\delta_{CP}\,, \\
\label{eq:invariants}&& I_{1}=\Im{(U^{2}_{12}U^{*\,2}_{13})}=\frac{1}{4}\sin^2\theta_{12}\sin^22\theta_{13}\sin(\beta+2\delta_{CP})\,.
\end{eqnarray}
For the mixing pattern in Eq.~\eqref{eq:PMNS_UGR1_Td}, these CP invariants turn out to be
\begin{equation}
\label{eq:invariants_GR_3col_Td}J_{CP}=\frac{\sin2\theta\sin\psi}{4 \sqrt{5(\phi+2)} }\,, \qquad I_{1}=\frac{1}{20\phi^2}\sin ^22 \theta  \sin (\rho -\sigma )\,.
\end{equation}
Since $J_{CP}$ and all the three mixing angles depend on only two parameters $\theta$ and $\psi$, we can derive the following sum rule among the Dirac CP phase $\delta_{CP}$ and mixing angles
\begin{equation}
\label{eq:sum_rule2_GR3}\cos\delta_{CP}=\frac{(\phi+2)  (1+\sin^ 2\theta_{13})-5 \cos^2\theta_{13}}{2 \sqrt{(\phi+2) (5 \cos ^2\theta_{13}- \phi-2 )}}\csc\theta_{13} \cot2\theta_{23}\,.
\end{equation}
For maximal atmospheric mixing angle $\theta_{23}=\pi/4$, this sum rule predicts $\cos\delta_{CP}=0$ which corresponds to maximal CP violation $\delta_{CP}=\pm\pi/2$. The mixing angles, CP phases and mass ratio $m_2/m_3$ depend on the $x$, $\eta$ and $r\equiv m_b/m_a$ while $m_2$ and $m_3$ depend on all the four input parameters $x$, $\eta$, $m_a$ and $m_b$. By comprehensively scanning over the parameter space of $\eta$ and $r$, we find that the experimental data on the mixing angles and mass squared splittings can be accommodated only for the values of $x=\pm2i\phi^2\sin\frac{2\pi}{5}$. In table~\ref{tab:best_fit_UGR1_3rd_col_Td} we present the predictions for the mixing angles and CP violation phases for some benchmark values of the parameters $\eta$ and $r$. It is remarkable that both atmospheric mixing angle and Dirac phase are maximal for $\eta=0$, all the mixing angles and mass ratio $m^2_2/m^2_3$ lie in the experimentally preferred $3\sigma$ ranges except that the reactor angle $\theta_{13}$ is a bit smaller. This tiny discrepancy is expected to be easily resolved in an explicit model with small corrections or by the renormalization group corrections~\cite{King:2016yef}. Notice that the same predictions for the mixing angles and maximal $\delta_{CP}$ can be obtained from the approach of combining $A_5$ flavor symmetry with generalized CP~\cite{Li:2015jxa,DiIura:2015kfa,Ballett:2015wia}, but we have additional prediction for the neutrino masses here even if the CP symmetry is not introduced in the present context. We can check that the neutrino mass matrix $m_{\nu}$ in Eq.~\eqref{eq:mnu_GR_3col_Td} has the following symmetry properties
\begin{eqnarray}
\nonumber&&m_{\nu}(\eta, x=\pm2i\phi^2\sin2\pi/5)=P^{T}_{23}m_{\nu}(\eta, x=\mp2i\phi^2\sin2\pi/5)P_{23},\\
\label{eq:sym_nu_matrix}&&m_{\nu}(\eta, x=\pm2i\phi^2\sin2\pi/5)=m^{*}_{\nu}(-\eta, x=\mp2i\phi^2\sin2\pi/5)\,,
\end{eqnarray}
with
\begin{equation}
P_{23}=\begin{pmatrix}
1 &~ 0 &~  0  \\
0 &~ 0  &~  1 \\
0 &~  1  &~  0
\end{pmatrix}\,.
\end{equation}
As a consequence, the same reactor and solar mixing angles are obtained for $x=2i\phi^2\sin\frac{2\pi}{5}$ and $x=-2i\phi^2\sin\frac{2\pi}{5}$, while the atmospheric
angle changes from $\theta_{23}$ to $\pi/2-\theta_{23}$ and the Dirac phase changes from $\delta_{CP}$ to $\pi+\delta_{CP}$. Moreover, all the lepton mixing angles are kept intact and the signs of all CP violation phases are reversed under the transformation $x\rightarrow -x$ and $\eta\rightarrow-\eta$. For the fixed value of $x=\pm2i\phi^2\sin\frac{2\pi}{5}$, all the mixing angles, CP phases and mass ratio $m^2_2/m^2_3$ are fully determined by $r$ and $\eta$, and the correct neutrino mass $m_2$ can be achieved for certain values of $m_b$. We show how these mixing parameters vary in the plane $\eta$ versus $r$ in figure~\ref{UGR1_3rd_col_Td}. It can be seen that the measured values of the mixing angles and the neutrino masses can be accommodated for certain choices of $\eta$ and $r$.

\begin{table}[hptb]
\renewcommand{\tabcolsep}{1.7mm}
\centering
\begin{tabular}{|c c c | c c c c c c |}  \hline \hline
\rule{0pt}{2.5ex}%
$\eta$    & $r$ & $x$  	 & $\sin^2\theta_{13}$  &$\sin^2\theta_{12}$  & $\sin^2\theta_{23}$  & $\delta_{CP}/\pi$ &  $\beta/\pi$  & $m^2_2/m^2_3$   \\ [0.5ex] \hline

\rule{0pt}{2.5ex} $0$ & $0.0177$ & $\pm 2i\phi^2\sin\frac{2\pi}{5}$ & $0.0164$ & $0.264$ & $0.5$ & $\mp0.5$ & $  0$ & $0.0309$ \\ [0.5ex] \hline
\rule{0pt}{2.5ex} $\pm\frac{\pi }{11}$ & $0.0185$ & $\pm 2i\phi^2\sin\frac{2\pi}{5}$ & $0.0174$ & $0.264$ & $0.614$ & $\mp0.331$ & $\mp0.210$ & $0.0302$ \\ [0.5ex] \hline
\rule{0pt}{2.5ex} $\pm\frac{\pi }{11}$ & $0.0185$ & $\mp 2i\phi^2\sin\frac{2\pi}{5}$ & $0.0175$ & $0.264$ & $0.385$ & $\pm0.670$ & $\mp0.211$ & $0.0304$ \\ [0.5ex] \hline
\rule{0pt}{2.5ex} $\pm\frac{\pi }{12}$ & $0.0183$ & $\pm 2i\phi^2\sin\frac{2\pi}{5}$ & $0.0172$ & $0.264$ & $0.605$ & $\mp0.345$ & $\mp0.192$ & $0.0303$ \\ [0.5ex] \hline
\rule{0pt}{2.5ex} $\pm\frac{\pi }{12}$ & $0.0184$ & $\mp 2i\phi^2\sin\frac{2\pi}{5}$ &  $0.0173$ & $0.264$ & $0.394$ & $\pm0.655$ & $\mp0.192$ & $0.0305$ \\ [0.5ex] \hline
\rule{0pt}{2.5ex} $\pm\frac{\pi }{13}$ & $0.0182$ & $\pm 2i\phi^2\sin\frac{2\pi}{5}$ & $0.0171$ & $0.264$ & $0.597$ & $\mp0.357$ & $\mp0.176$ & $0.0304$ \\ [0.5ex] \hline
\rule{0pt}{2.5ex} $\pm\frac{\pi }{13}$ & $0.0183$ & $\mp 2i\phi^2\sin\frac{2\pi}{5}$ &  $0.0172$ & $0.264$ & $0.402$ & $\pm0.643$ & $\mp0.177$ & $0.0306$ \\ [0.5ex] \hline
\rule{0pt}{2.5ex} $\pm\frac{\pi }{14}$ & $0.0182$ & $\pm 2i\phi^2\sin\frac{2\pi}{5}$ &  $0.0170$ & $0.264$ & $0.591$ & $\mp0.368$ & $\mp0.163$ & $0.0304$ \\ [0.5ex] \hline
\rule{0pt}{2.5ex} $\pm\frac{\pi }{14}$ & $0.0182$ & $\mp 2i\phi^2\sin\frac{2\pi}{5}$ &  $0.0171$ & $0.264$ & $0.409$ & $\pm0.632$ & $\mp0.164$ & $0.0307$ \\ [0.5ex] \hline
\rule{0pt}{2.5ex} $\pm\frac{\pi }{15}$ & $0.0181$ & $\pm 2i\phi^2\sin\frac{2\pi}{5}$ &  $0.0169$ & $0.264$ & $0.585$ & $\mp0.377$ & $\mp0.152$ & $0.0305$ \\ [0.5ex] \hline
\rule{0pt}{2.5ex} $\pm\frac{\pi }{15}$ & $0.0181$ & $\mp 2i\phi^2\sin\frac{2\pi}{5}$ & $0.0170$ & $0.264$ & $0.415$ & $\pm0.623$ & $\mp0.152$ & $0.0307$ \\ [0.5ex] \hline
\rule{0pt}{2.5ex} $\pm\frac{\pi }{16}$ & $0.0180$ & $\pm 2i\phi^2\sin\frac{2\pi}{5}$ &  $0.0168$ & $0.264$ & $0.580$ & $\mp0.385$ & $\mp0.142$ & $0.0305$ \\ [0.5ex] \hline
\rule{0pt}{2.5ex} $\pm\frac{\pi }{16}$ & $0.0181$ & $\mp 2i\phi^2\sin\frac{2\pi}{5}$ & $0.0169$ & $0.264$ & $0.420$ & $\pm0.616$ & $\mp0.142$ & $0.0308$ \\ [0.5ex] \hline
\rule{0pt}{2.5ex} $\pm\frac{\pi }{17}$ & $0.0180$ & $\pm 2i\phi^2\sin\frac{2\pi}{5}$ & $0.0168$ & $0.264$ & $0.575$ & $\mp0.391$ & $\mp0.134$ & $0.0306$ \\ [0.5ex] \hline
\rule{0pt}{2.5ex} $\pm\frac{\pi }{17}$ & $0.0180$ & $\mp 2i\phi^2\sin\frac{2\pi}{5}$ & $0.0169$ & $0.264$ & $0.425$ & $\pm0.609$ & $\mp0.134$ & $0.0308$ \\ [0.5ex] \hline
\rule{0pt}{2.5ex} $\pm\frac{\pi }{18}$ & $0.0180$ & $\pm 2i\phi^2\sin\frac{2\pi}{5}$ & $0.0167$ & $0.264$ & $0.571$ & $\mp0.398$ & $\mp0.126$ & $0.0306$ \\ [0.5ex] \hline
\rule{0pt}{2.5ex} $\pm\frac{\pi }{18}$ & $0.0180$ & $\mp 2i\phi^2\sin\frac{2\pi}{5}$ & $0.0168$ & $0.264$ & $0.429$ & $\pm0.603$ & $\mp0.126$ & $0.0308$ \\ [0.5ex] \hline
\rule{0pt}{2.5ex} $\pm\frac{\pi }{19}$ & $0.0179$ & $\pm 2i\phi^2\sin\frac{2\pi}{5}$ & $0.0167$ & $0.264$ & $0.567$ & $\mp0.403$ & $\mp0.119$ & $0.0306$ \\ [0.5ex] \hline
\rule{0pt}{2.5ex} $\pm\frac{\pi }{19}$ & $0.0180$ & $\mp 2i\phi^2\sin\frac{2\pi}{5}$ & $0.0168$ & $0.264$ & $0.432$ & $\pm0.597$ & $\mp0.119$ & $0.0308$ \\ [0.5ex] \hline
\rule{0pt}{2.5ex} $\pm\frac{\pi }{20}$ & $0.0179$ & $\pm 2i\phi^2\sin\frac{2\pi}{5}$ & $0.0167$ & $0.264$ & $0.564$ & $\mp0.408$ & $\mp0.113$ & $0.0306$ \\ [0.5ex] \hline
\rule{0pt}{2.5ex} $\pm\frac{\pi }{20}$ & $0.0179$ & $\mp 2i\phi^2\sin\frac{2\pi}{5}$ & $0.0167$ & $0.264$ & $0.436$ & $\pm0.592$ & $\mp0.113$ & $0.0308$ \\ [0.5ex] \hline
\rule{0pt}{2.5ex} $\pm\frac{\pi }{21}$ & $0.0179$ & $\pm 2i\phi^2\sin\frac{2\pi}{5}$ & $0.0166$ & $0.264$ & $0.561$ & $\mp0.412$ & $\mp0.108$ & $0.0306$ \\ [0.5ex] \hline
\rule{0pt}{2.5ex} $\pm\frac{\pi }{21}$ & $0.0179$ & $\mp 2i\phi^2\sin\frac{2\pi}{5}$ & $0.0167$ & $0.264$ & $0.439$ & $\pm0.588$ & $\mp0.108$ & $0.0308$ \\ [0.5ex] \hline
\rule{0pt}{2.5ex} $\pm\frac{\pi }{22}$ & $0.0179$ & $\pm 2i\phi^2\sin\frac{2\pi}{5}$ & $0.0166$ & $0.264$ & $0.558$ & $\mp0.416$ & $\mp0.103$ & $0.0307$ \\ [0.5ex] \hline
\rule{0pt}{2.5ex} $\pm\frac{\pi }{22}$ & $0.0179$ & $\mp 2i\phi^2\sin\frac{2\pi}{5}$ & $0.0167$ & $0.264$ & $0.442$ & $\pm0.584$ & $\mp0.103$ & $0.0309$ \\ [0.5ex] \hline
\rule{0pt}{2.5ex} $\pm\frac{\pi }{23}$ & $0.0179$ & $\pm 2i\phi^2\sin\frac{2\pi}{5}$ &  $0.0166$ & $0.264$ & $0.556$ & $\mp0.420$ & $\mp0.0982$ & $0.0307$ \\ [0.5ex] \hline
\rule{0pt}{2.5ex} $\pm\frac{\pi }{23}$ & $0.0179$ & $\mp 2i\phi^2\sin\frac{2\pi}{5}$ & $0.0167$ & $0.264$ & $0.444$ & $\pm0.580$ & $\mp0.0983$ & $0.0309$ \\ [0.5ex] \hline
\rule{0pt}{2.5ex} $\pm\frac{2 \pi }{23}$ & $0.0184$ & $\pm 2i\phi^2\sin\frac{2\pi}{5}$ & $0.0173$ & $0.264$ & $0.610$ & $\mp0.338$ & $\mp0.201$ & $0.0302$ \\ [0.5ex] \hline
\rule{0pt}{2.5ex} $\pm\frac{2 \pi }{23}$ & $0.0184$ & $\mp 2i\phi^2\sin\frac{2\pi}{5}$ &  $0.0174$ & $0.264$ & $0.390$ & $\pm0.662$ & $\mp0.201$ & $0.0305$ \\ [0.5ex] \hline\hline
\end{tabular}
\caption{\label{tab:best_fit_UGR1_3rd_col_Td}Predictions for all the lepton mixing angles, CP violation phases and $m^2_2/m^2_3$ in the golden Littlest seesaw with $\Phi_{\mathrm{atm}}\propto\Phi_{3}$. Here we choose many benchmark values for the parameters $\eta$ and $r$.}
\end{table}

\subsubsection{\label{subsubsec:UGR1_2nd_col_Td}Golden Littlest seesaw with $\Phi_{\mathrm{atm}}\propto\Phi_{2}$ }

\begin{figure}[t!]
\centering
\begin{tabular}{c}
\includegraphics[width=0.99\linewidth]{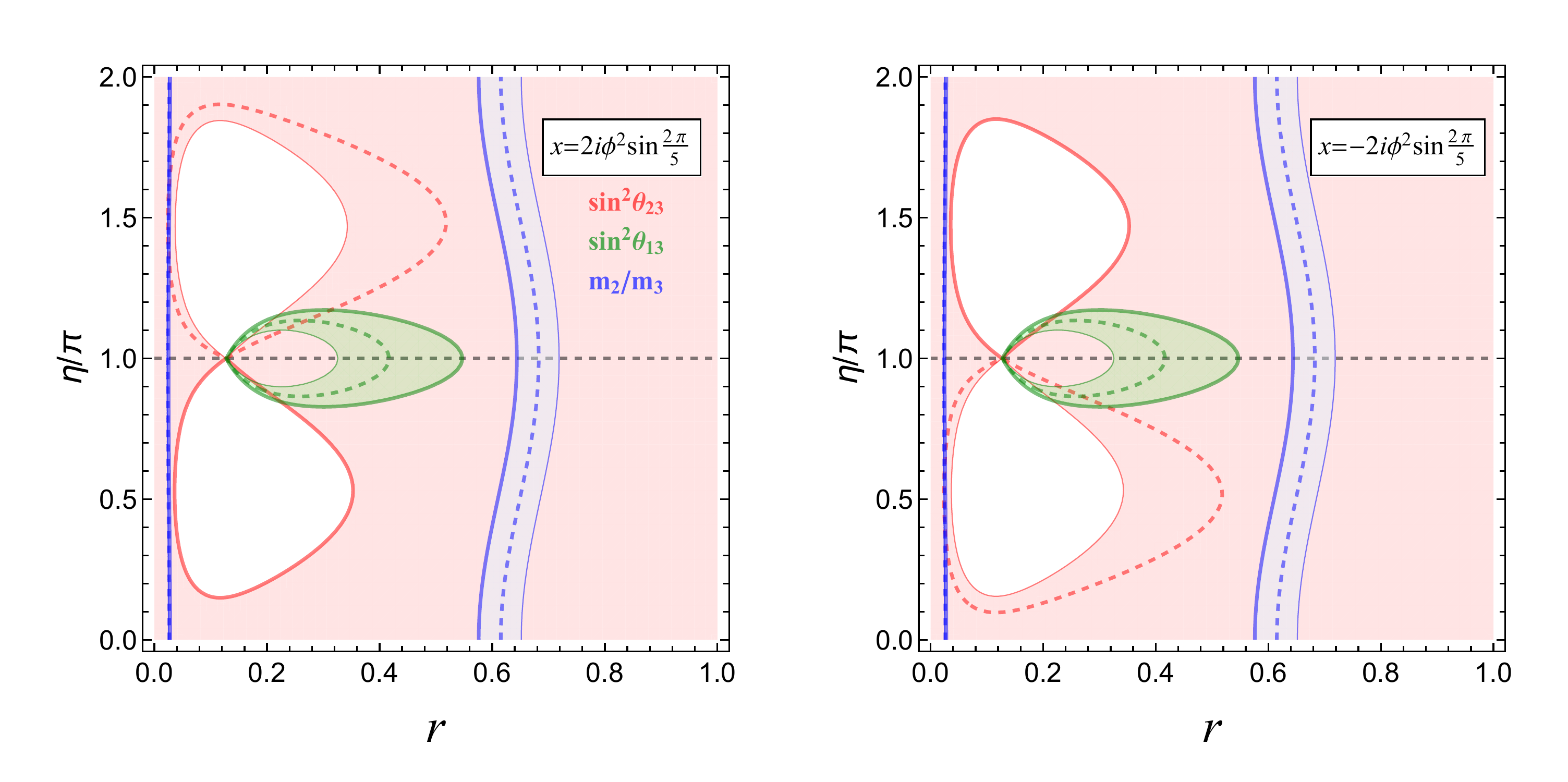}
\end{tabular}
\caption{\label{UGR1_2nd_col_Td} Contour plots of $\sin^2\theta_{13}$, $\sin^2\theta_{23}$ and $m_2/m_3$ in the $\eta-r$ plane for the golden Littlest seesaw with $\Phi_{\mathrm{atm}}\propto\Phi_{2}$. Here we take $x=2i\phi^2\sin(2\pi/5)$ and $x=-2i\phi^2\sin(2\pi/5)$ for which the solar vacuum alignment $\Phi_{\text{sol}}$ preserves the residual symmetry $G_{\text{sol}}=Z^{T^3ST^2S}_3$ and $G_{\text{sol}}=Z^{ST^2ST^3}_3$ respectively. The $3\sigma$ upper (lower) bounds of the lepton mixing angles are labelled with thick (thin) solid curves, and the dashed contour lines represent the corresponding best fit values. The $3\sigma$ ranges as well as the best fit values of the mixing angles are adapted from~\cite{Capozzi:2017ipn}. The black contour line refers to maximal atmospheric mixing angle with $\sin^2\theta_{23}=0.5$.}
\end{figure}

Similar to previous case, the most general from of the solar vacuum $\Phi_{\text{sol}}$ is given by Eq.~\eqref{eq:phi_sol}, and the atmospheric alignment vector takes the form
\begin{equation}
\Phi_{\text{atm}}\propto \left(\sqrt{2},~\phi,~\phi\right)^{T}\,,
\end{equation}
which preserves the residual symmetry $G_{\text{atm}}=Z^{T^3ST^2ST^3}_2$. Subsequently we can read out the Dirac neutrino mass matrix $M_D$ and the right-handed neutrino mass matrix $M_{N}$ as
\begin{equation}
M_{D}=\begin{pmatrix}
\sqrt{2}a ~&~   \sqrt{2} b \\
\phi a   ~&~  (\phi-x) b \\
\phi a  ~&~  (\phi +x)b
\end{pmatrix}\,,\qquad M_{N}=\begin{pmatrix}
M_{\mathrm{atm}}  &  0  \\
0  &   M_{\mathrm{sol}}
\end{pmatrix}\,,
\end{equation}
which leads to the following low energy effective Majorana neutrino mass matrix
\begin{equation}
\label{eq:mnu_GR_2col_Td}m_{\nu}=m_a\begin{pmatrix}
 2 &~ \sqrt{2} \phi  ~& \sqrt{2} \phi  \\
 \sqrt{2} \phi  &~ \phi +1 ~& \phi +1 \\
 \sqrt{2} \phi  &~ \phi +1 ~& \phi +1 \\
\end{pmatrix}+m_{b}\,e^{i\eta}
\begin{pmatrix}
 2 &~ \sqrt{2} (\phi -x) ~& \sqrt{2} (x+\phi ) \\
 \sqrt{2} (\phi -x) &~ (x-\phi )^2 ~& -x^2+\phi +1 \\
 \sqrt{2} (x+\phi ) &~ -x^2+\phi +1 ~& (x+\phi )^2 \\
\end{pmatrix}\,,
\end{equation}
with $m_a=|a|^2/M_{\mathrm{atm}}$, $m_b=|b|^2/M_{\mathrm{sol}}$ and $\eta=\mathrm{arg}(b^2/a^2)$. This model is rather predictive since only four parameters $m_a$, $m_{b}$, $x$ and $\eta$ can describe the entire neutrino sector. The symmetry relations in Eq.~\eqref{eq:sym_nu_matrix} are also satisfied in this case. The neutrino mass matrix in Eq.~\eqref{eq:mnu_GR_2col_Td} can be block diagonalized by the GR mixing matrix,
\begin{equation}
\label{eq:mnu_UGR2_Td}
m^{\prime}_{\nu}=U^{T}_{GR}m_{\nu}U_{GR}=\begin{pmatrix}
0  &~   0  &~  0 \\
0  &~  y   &~ z  \\
0  &~  z  &~ w
\end{pmatrix}\,,
\end{equation}
where
\begin{eqnarray}
\nonumber&&y=|y|e^{i\phi_{y}}=2\sqrt{5}\,\phi \left(m_a+m_{b}\,e^{i\eta}\right),\\
\nonumber&&z=2x\sqrt{\phi+2}\,m_{b}\,e^{i\eta},\\
\label{eq:xyz_UGR_2col_Td}&&w=2x^2\,m_{b}\,e^{i\eta}\,.
\end{eqnarray}
Furthermore, $m'_{\nu}$ can be put into diagonal form by performing another unitary transformation
\begin{equation}
{U^{\prime}}^{T}m^{\prime}_{\nu}U^{\prime}=\text{diag}(0, m_2, m_3)\,,
\end{equation}
with
\begin{equation}
\label{eq:Uprime}U^{\prime}=\begin{pmatrix}
1  &~    0    &~   0 \\
0  &~  \cos\theta \,e^{i(\psi+\rho)/2}   &~  \sin\theta \,e^{i(\psi+\sigma)/2}   \\
0  &~   -\sin\theta\, e^{i(-\psi+\rho)/2}    &~~    \cos\theta\, e^{i(-\psi+\sigma)/2}
\end{pmatrix}\,,
\end{equation}
where the parameters $\theta$,  $\psi$, $\rho$ and $\sigma$ are determined in terms of $x$, $y$, $z$ defined in Eq.~\eqref{eq:xyz_UGR_2col_Td},
\begin{eqnarray}
\nonumber&&\sin2\theta=\frac{-2iz\,e^{-i\eta}\sqrt{|y|^2+|w|^2-2|y||w|\cos(\phi_{y}-\eta)}}{\sqrt{(|w|^2-|y|^2)^2+4|z|^2\left[|y|^2+|w|^2-2|y||w|\cos(\phi_{y}-\eta)\right]}},\\
\nonumber&&\cos2\theta=\frac{|w|^2-|y|^2}{\sqrt{(|w|^2-|y|^2)^2+4|z|^2\left[|y|^2+|w|^2-2|y||w|\cos(\phi_{y}-\eta)\right]}}\,,\\
\nonumber&&\sin\psi=\frac{|y|\cos(\phi_{y}-\eta)-|w|}{\sqrt{|y|^2+|w|^2-2|y||w|\cos(\phi_{y}-\eta)}}\,,\\
\nonumber&&\cos\psi=\frac{|y|\sin(\phi_{y}-\eta)}{\sqrt{|y|^2+|w|^2-2|y||w|\cos(\phi_{y}-\eta)}},\\
\nonumber&&\sin\rho=-\frac{(m^2_2-|z|^2)\cos\eta-|y||w|\cos\phi_{y}}{m_2\sqrt{|y|^2+|w|^2-2|y||w|\cos(\phi_{y}-\eta)}}\,,\\
\nonumber&&\cos\rho=\frac{-(m^2_2-|z|^2)\sin\eta+|y||w|\sin\phi_{y}}{m_2\sqrt{|y|^2+|w|^2-2|y||w|\cos(\phi_{y}-\eta)}}\,,\\
\nonumber&&\sin\sigma=-\frac{(m^2_3-|z|^2)\cos\eta-|y||w|\cos\phi_y}{m_3\sqrt{|y|^2+|w|^2-2|y||w|\cos(\phi_{y}-\eta)}}\,,\\
&&\cos\sigma=\frac{-(m^2_3-|z|^2)\sin\eta+|y||w|\sin\phi_y}{m_3\sqrt{|y|^2+|w|^2-2|y||w|\cos(\phi_{y}-\eta)}}\,.
\end{eqnarray}
The exact expressions for the neutrino masses are given by
\begin{eqnarray}
\nonumber&&m^2_1=0\,,\\
\nonumber&&m^2_2=\frac{1}{2}\left[|y|^2+|w|^2+2|z|^2-\frac{|w|^2-|y|^2}{\cos2\theta}\right]\,,\\
\label{eq:nu_mass_UGR1}&&m^2_3=\frac{1}{2}\left[|y|^2+|w|^2+2|z|^2+\frac{|w|^2-|y|^2}{\cos2\theta}\right]
\end{eqnarray}
Given that the charged lepton mass matrix is diagonal due to the $Z^{T}_5$ residual symmetry, the PMNS mixing matrix is of the form
\begin{equation}\label{eq:PMNS_UGR1_Benz}
\hskip-0.1in U=U_{GR}U^{\prime}=\sqrt{\frac{\phi-1}{2\sqrt{5}}}
\begin{pmatrix}
 -\sqrt{2} \phi  &~ \sqrt{2} \cos \theta  ~& \sqrt{2} e^{i \psi } \sin \theta  \\
 1 &~ \phi  \cos \theta +\sqrt{\phi+2}  \sin \theta\, e^{-i \psi }  ~&  \phi  \sin \theta \,e^{i \psi } -\sqrt{\phi+2}  \cos \theta  \\
 1 &~ \phi  \cos \theta -\sqrt{\phi+2} \sin \theta \, e^{-i \psi } ~&  \phi  \sin \theta \,e^{i \psi } +\sqrt{\phi+2} \cos \theta   \\
\end{pmatrix}P_{\nu}\,,
\end{equation}
with
\begin{equation}
P_{\nu}=\text{diag}(1, e^{i(\psi+\rho)/2}, e^{i(-\psi+\sigma)/2})\,.
\end{equation}
Obviously the first column of the mixing matrix is fixed to be that of the GR mixing matrix. The lepton mixing matrix $U$ is identical with the one in Eq.~\eqref{eq:PMNS_UGR_3col_Td}. Hence all the mixing angles and CP invariants are predicted to have the same form as those of Eq.~\eqref{eq:angles_GR_3col_Td} and Eq.~\eqref{eq:invariants_GR_3col_Td} respectively, but their dependence on the input parameters $m_a$, $m_b$, $\eta$ and $x$ are different. The sum rules in Eq.~\eqref{eq:sum_rule_GR3} and Eq.~\eqref{eq:sum_rule2_GR3} are satisfied as well. Detailed numerical analyses show that accordance with experimental data can be achieved for certain values of $r=m_b/m_a$ and $\eta$ in the case of $x=\pm2i\phi^2\sin\frac{2\pi}{5}$, and the corresponding benchmark numerical results are listed in table~\ref{tab:best_fit_UGR1_2nd_col_Td}. The most interesting point is $\eta=\pi$ which predicts maximal atmospheric mixing and a maximal Dirac phase. The realistic values of $\sin^2\theta_{12}$
and $m^2_2/m^2_3$ can be obtained for $r=1.486$ while the reactor angle is slightly a bit larger. This mixing pattern for $\eta=\pi$ can also be obtained from $A_5$ flavor symmetry and CP in the semidirect approach~\cite{Li:2015jxa,DiIura:2015kfa,Ballett:2015wia}, the additional bonus in GLS is the predcition for neutrino masses. As discussed in above, all the mixing parameters as well as mass ratio $m_2/m_3$ depend only on $\eta$ and $r$, this dependence is shown in figure~\ref{UGR1_2nd_col_Td}.

\begin{figure}[hptb!]
\centering
\begin{tabular}{c}
\includegraphics[width=0.97\linewidth]{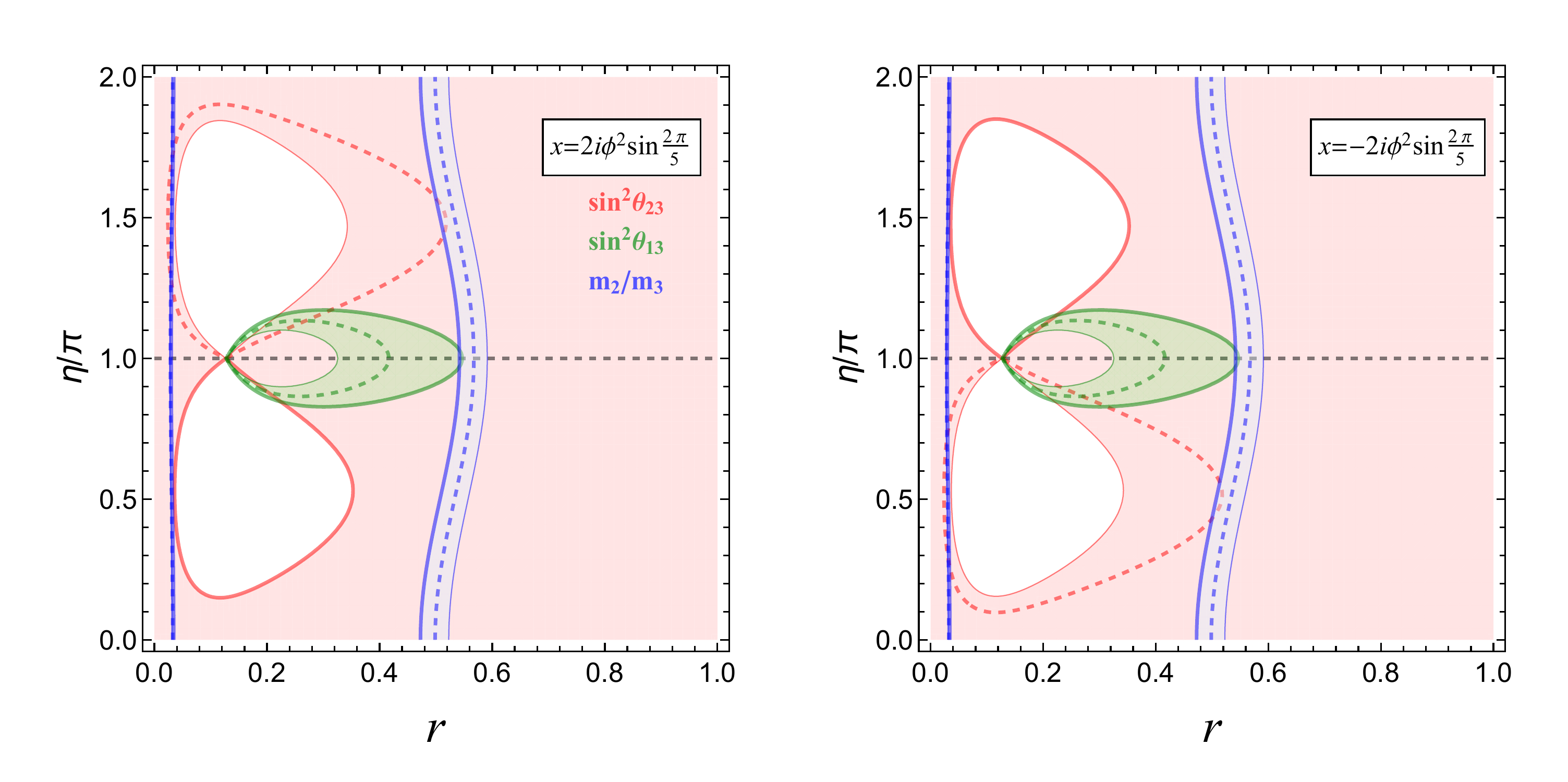}
\end{tabular}
\caption{\label{UGR1_2nd_col_Td_Dec}Contour plots of $\sin^2\theta_{13}$, $\sin^2\theta_{23}$ and $m_2/m_3$ in the $\eta-r$ plane for the golden Littlest seesaw with $\Phi_{\mathrm{atm}}\propto\Phi_{2}$. As an example,  we assume that the decoupled alignment $\Phi_{\text{dec}}\propto\Phi_1$ which gives rise to $m_{1}=6\times10^{-3}$eV.}
\end{figure}

If we further take into account the contribution of the third almost decoupled right-handed neutrino of mass $M_{\text{dec}}$, for example for the case of $\Phi_{\text{dec}}\propto\Phi_1$, the last term of Eq.~\eqref{eq:mnu_indirect} would contribute to the lightest neutrino mass $m_1=c^2/M_{\text{dec}}$, while the neutrino mixing angles, CP violating phases and the other two neutrino masses are not changed. From figure~\ref{UGR1_2nd_col_Td_Dec}, we can see that better agreement with experimental data can be achieved. The viable regions for $\sin^2\theta_{13}$, $\sin^2\theta_{23}$ and $m_2/m_3$ can overlap with each other.

\begin{table}[hptb]
\renewcommand{\tabcolsep}{1.7mm}
\centering
\begin{tabular}{|c c c | c c c c c c |}  \hline \hline
\rule{0pt}{2.5ex}%
$\eta$   & $r$ & $x$
  	 &  $\sin^2\theta_{13}$  &$\sin^2\theta_{12}$  & $\sin^2\theta_{23}$  & $\delta_{CP}/\pi$ &  $\beta/\pi$  & $m^2_2/m^2_3$   \\ [0.5ex] \hline

\rule{0pt}{2.5ex} $\pi$ & $0.675$ & $\pm2i\phi^2\sin\frac{2\pi}{5}$ & $0.0257$ & $0.257$ & $\pm0.5$ & $0.5$ & $0$ & $0.0294$ \\ [0.5ex] \hline
\rule{0pt}{2.5ex} $\pm\frac{4 \pi }{5}$ & $0.670$ & $\pm2i\phi^2\sin\frac{2\pi}{5}$ &  $0.0282$ & $0.255$ & $0.535$ & $\pm0.465$ & $\mp0.203$ & $0.0293$ \\ [0.5ex] \hline
\rule{0pt}{2.5ex} $\pm\frac{4 \pi }{5}$ & $0.669$ & $\mp2i\phi^2\sin\frac{2\pi}{5}$ & $0.0282$ & $0.255$ & $0.465$ & $\mp0.536$ & $\mp0.203$ & $0.0294$ \\ [0.5ex] \hline
\rule{0pt}{2.5ex} $\pm\frac{5 \pi }{6}$ & $0.671$ & $\pm2i\phi^2\sin\frac{2\pi}{5}$ & $0.0275$ & $0.256$ & $0.529$ & $\pm0.469$ & $\mp0.169$ & $0.0294$ \\ [0.5ex] \hline
\rule{0pt}{2.5ex} $\pm\frac{5 \pi }{6}$ & $0.67$ & $\mp2i\phi^2\sin\frac{2\pi}{5}$ &  $0.0274$ & $0.256$ & $0.47$ & $\mp0.531$ & $\mp0.169$ & $0.0295$ \\ [0.5ex] \hline
\rule{0pt}{2.5ex} $\pm\frac{6 \pi }{7}$ & $0.672$ & $\pm2i\phi^2\sin\frac{2\pi}{5}$ & $0.027$ & $0.256$ & $0.526$ & $\pm0.473$ & $\mp0.145$ & $0.0294$ \\ [0.5ex] \hline
\rule{0pt}{2.5ex} $\pm\frac{6 \pi }{7}$ & $0.671$ & $\mp2i\phi^2\sin\frac{2\pi}{5}$ & $0.027$ & $0.256$ & $0.474$ & $\mp0.527$ & $\mp0.145$ & $0.0295$ \\ [0.5ex] \hline
\rule{0pt}{2.5ex} $\pm\frac{7 \pi }{8}$ & $0.673$ & $\pm2i\phi^2\sin\frac{2\pi}{5}$ & $0.0267$ & $0.257$ & $0.523$ & $\pm0.476$ & $\mp0.127$ & $0.0294$ \\ [0.5ex] \hline
\rule{0pt}{2.5ex} $\pm\frac{7 \pi }{8}$ & $0.672$ & $\mp2i\phi^2\sin\frac{2\pi}{5}$ & $0.0267$ & $0.257$ & $0.477$ & $\mp0.524$ & $\mp0.127$ & $0.0295$ \\ [0.5ex] \hline
\rule{0pt}{2.5ex} $\pm\frac{8 \pi }{9}$ & $0.674$ & $\pm2i\phi^2\sin\frac{2\pi}{5}$ & $0.0265$ & $0.257$ & $0.52$ & $\pm0.479$ & $\mp0.113$ & $0.0294$ \\ [0.5ex] \hline
\rule{0pt}{2.5ex} $\pm\frac{8 \pi }{9}$ & $0.673$ & $\mp2i\phi^2\sin\frac{2\pi}{5}$ & $0.0265$ & $0.257$ & $0.48$ & $\mp0.521$ & $\mp0.113$ & $0.0295$ \\ [0.5ex] \hline
\rule{0pt}{2.5ex} $\pm\frac{9 \pi }{10}$ & $0.674$ & $\pm2i\phi^2\sin\frac{2\pi}{5}$ & $0.0264$ & $0.257$ & $0.518$ & $\pm0.481$ & $\mp0.101$ & $0.0294$ \\ [0.5ex] \hline
\rule{0pt}{2.5ex} $\pm\frac{9 \pi }{10}$ & $0.673$ & $\mp2i\phi^2\sin\frac{2\pi}{5}$ &  $0.0263$ & $0.257$ & $0.482$ & $\mp0.519$ & $\mp0.101$ & $0.0295$ \\ [0.5ex] \hline
\rule{0pt}{2.5ex} $\pm\frac{10 \pi }{11}$ & $0.674$ & $\pm2i\phi^2\sin\frac{2\pi}{5}$ & $0.0262$ & $0.257$ & $0.517$ & $\pm0.482$ & $\mp0.0922$ & $0.0294$ \\ [0.5ex] \hline
\rule{0pt}{2.5ex} $\pm\frac{10 \pi }{11}$ & $0.674$ & $\mp2i\phi^2\sin\frac{2\pi}{5}$ &  $0.0262$ & $0.257$ & $0.483$ & $\mp0.518$ & $\mp0.0922$ & $0.0294$ \\ [0.5ex] \hline
\rule{0pt}{2.5ex} $\pm\frac{11 \pi }{12}$ & $0.674$ & $\pm2i\phi^2\sin\frac{2\pi}{5}$ &  $0.0262$ & $0.257$ & $0.515$ & $\pm0.484$ & $\mp0.0845$ & $0.0294$ \\ [0.5ex] \hline
\rule{0pt}{2.5ex} $\pm\frac{11 \pi }{12}$ & $0.674$ & $\mp2i\phi^2\sin\frac{2\pi}{5}$ &  $0.0262$ & $0.257$ & $0.485$ & $\mp0.516$ & $\mp0.0845$ & $0.0294$ \\ [0.5ex] \hline
\rule{0pt}{2.5ex} $\pm\frac{12 \pi }{13}$ & $0.675$ & $\pm2i\phi^2\sin\frac{2\pi}{5}$ & $0.0261$ & $0.257$ & $0.514$ & $\pm0.485$ & $\mp0.078$ & $0.0294$ \\ [0.5ex] \hline
\rule{0pt}{2.5ex} $\pm\frac{12 \pi }{13}$ & $0.674$ & $\mp2i\phi^2\sin\frac{2\pi}{5}$  & $0.0261$ & $0.257$ & $0.486$ & $\mp0.515$ & $\mp0.078$ & $0.0294$ \\ [0.5ex] \hline
\rule{0pt}{2.5ex} $\pm\frac{13 \pi }{14}$ & $0.675$ & $\pm2i\phi^2\sin\frac{2\pi}{5}$ & $0.0260$ & $0.257$ & $0.513$ & $\pm0.486$ & $\mp0.0724$ & $0.0294$ \\ [0.5ex] \hline
\rule{0pt}{2.5ex} $\pm\frac{13 \pi }{14}$ & $0.674$ & $\mp2i\phi^2\sin\frac{2\pi}{5}$ &  $0.0260$ & $0.257$ & $0.487$ & $\mp0.514$ & $\mp0.0724$ & $0.0294$ \\ [0.5ex] \hline
\rule{0pt}{2.5ex} $\pm\frac{13 \pi }{15}$ & $0.673$ & $\pm2i\phi^2\sin\frac{2\pi}{5}$ &  $0.0268$ & $0.256$ & $0.524$ & $\pm0.475$ & $\mp0.135$ & $0.0294$ \\ [0.5ex] \hline
\rule{0pt}{2.5ex} $\pm\frac{13 \pi }{15}$ & $0.672$ & $\mp2i\phi^2\sin\frac{2\pi}{5}$ & $0.0268$ & $0.256$ & $0.476$ & $\mp0.525$ & $\mp0.135$ & $0.0295$ \\ [0.5ex] \hline
\rule{0pt}{2.5ex} $\pm\frac{14 \pi }{15}$ & $0.675$ & $\pm2i\phi^2\sin\frac{2\pi}{5}$ & $0.0260$ & $0.257$ & $0.512$ & $\pm0.487$ & $\mp0.0676$ & $0.0294$ \\ [0.5ex] \hline
\rule{0pt}{2.5ex} $\pm\frac{14 \pi }{15}$ & $0.674$ & $\mp2i\phi^2\sin\frac{2\pi}{5}$ &  $0.0260$ & $0.257$ & $0.488$ & $\mp0.513$ & $\mp0.0676$ & $0.0294$ \\ [0.5ex] \hline \hline
\end{tabular}
\caption{\label{tab:best_fit_UGR1_2nd_col_Td}
Benchmark numerical results in the golden Littlest seesaw for the case of $\Phi_{\mathrm{atm}}\propto\Phi_{2}$ and $x=\pm2i\phi^2\sin(2\pi/5)$. }
\end{table}

\newpage

\section{\label{sec:possible_LS}Alternative Golden Littlest Seesaw in $A_5$}

In the direct approach, if the $A_5$ flavor symmetry is broken down to Klein subgroups in both the neutrino and charged lepton sectors, e.g. $G_{l}=K^{(S,T^3ST^2ST^3)}_4$ and $G_{\nu}=K^{(ST^2ST^3S, TST^4)}_4$, the lepton mixing matrix is determined to be of the row-column ($RC$) symmetric form~\cite{Li:2015jxa,deAdelhartToorop:2011re}
\begin{equation}
\label{eq:URC}
U_{RC}=\frac{1}{2}\begin{pmatrix}
\phi   &~  -1   &~  1/\phi \\
-1   &~  -1/\phi   &~   \phi  \\
1/\phi   &~  \phi   &~  1
\end{pmatrix} \,.
\end{equation}
The mixing angles are: $\sin^2\theta_{12}=\left(3-\phi\right)/5\simeq0.276$,  $\sin^2\theta_{23}=\left(2+\phi\right)/5\simeq0.724$ and  $\sin^2\theta_{13}=\left(2-\phi\right)/4\simeq0.0955$.
Although this mixing pattern is not phenomenologically viable because of too large $\theta_{13}$ and $\theta_{23}$, the first column of $U_{RC}$ is still compatible with experimental data, and that is what we shall assume in the following.

\begin{figure}[t!]
\centering
\begin{tabular}{c}
\includegraphics[width=0.50\linewidth]{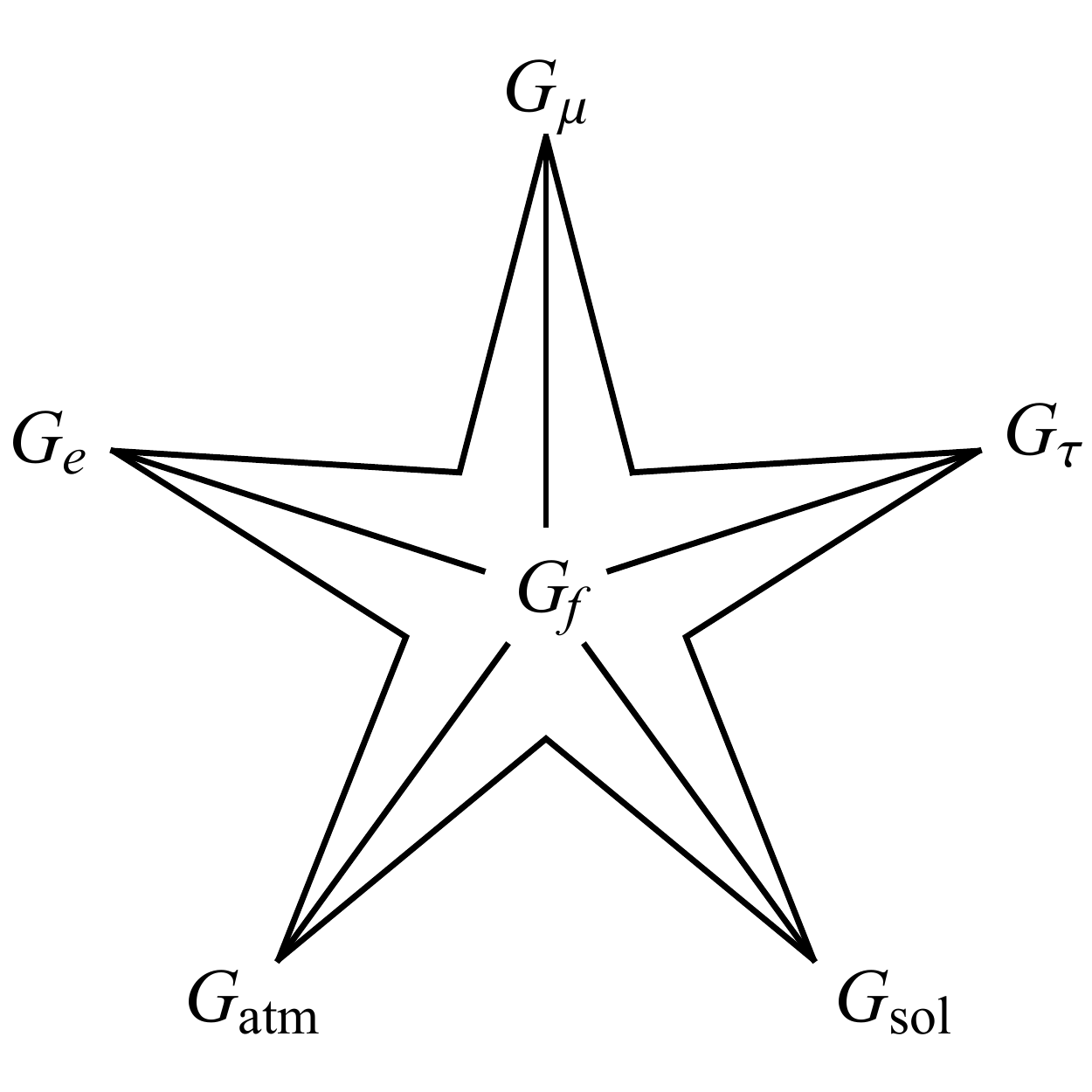}
\end{tabular}
\caption{\label{fig:star} A sketch of the indirect model building approach, where the electron, muon and tau sectors preserve different residual subgroups $G_{e}$, $G_{\mu}$ and $G_{\tau}$ respectively, and the neutrino vacuum alignments $\Phi_{\text{atm}}$ and $\Phi_{\text{sol}}$ are enforced by the residual symmetries $G_{\text{atm}}$ and $G_{\text{sol}}$ respectively.}
\end{figure}

The general principle of the Littlest seesaw is that different sectors of the Lagrangian preserve different residual subgroups of the flavor symmetry which is proposed in Ref.~\cite{King:2015dvf}. In this section, we shall consider the case that the electron, muon and tau sectors preserve different residual symmetries while the flavor symmetry is broken in the whole charged lepton Lagrangian, and the same holds true for the neutrino vacuum $\Phi_{\text{atm}}$ and $\Phi_{\text{sol}}$. This scenario is schematically depicted in figure~\ref{fig:star}. Moreover we can generally write down the Littlest seesaw Lagrangian in the neutrino and charged lepton sectors as follows:
\begin{eqnarray}
\nonumber\mathcal{L}&=&-y_{\mathrm{atm}}\bar{L}.\phi_{\mathrm{atm}}N_R^{\mathrm{atm}}-y_{\mathrm{sol}}\bar{L}.\phi_{\mathrm{sol}}N_R^{\mathrm{sol}}
-\frac{1}{2}M_{\mathrm{atm}}\overline{(N^{\mathrm{atm}}_R)^c}N_R^{\mathrm{atm}}-\frac{1}{2}M_{\mathrm{sol}}\overline{(N^{\mathrm{sol}}_R)^c}N_R^{\mathrm{sol}}\\
\label{eq:Lagrangian_star}&&+y_{\tau}\bar{L}.\varphi_{\tau}\tau_{R}+y_{\mu}\bar{L}.\varphi_{\mu}\mu_{R}+y_{e}\bar{L}.\varphi_{e}e_{R}+{h.c.}\,,
\end{eqnarray}
where $\varphi_{\alpha}$ ($\alpha=e, \mu, \tau$) can be Higgs fields transforming as triplets under the flavor symmetry group, or the combination of the electroweak Higgs doublet with triplet scalar flavons. In order to obtain the above terms in a concrete model, the possible additional abelian symmetries are generically needed and they will not be specified here. It is generally more convenient to work in the charged lepton diagonal basis in practical model building. We show such an appropriate alternative basis in table~\ref{tab:representation}.
In this basis the charged lepton mass matrix is enforced to be diagonal by the chosen residual symmetries $G_{e}$, $G_{\mu}$ and $G_{\tau}$ in Eq.~\eqref{eq:G_e_mu_tau}. The desired vacuum alignments in the charged lepton sector are
\begin{equation}
\label{eq:vacuum_clepton}\langle \varphi_e \rangle =v_e
\begin{pmatrix} 1\\0\\0 \end{pmatrix}  \ , \qquad
\langle \varphi_\mu \rangle =v_\mu
\begin{pmatrix} 0\\1\\0 \end{pmatrix} \ , \qquad
\langle \varphi_\tau \rangle = v_\tau
\begin{pmatrix} 0\\0\\1 \end{pmatrix}
\,.
\end{equation}
If we regard $\varphi_e$, $\varphi_{\mu}$, $ \varphi_{\tau}$ as each being a triplet $\mathbf{3}$ of $A_5$, then they each correspond to a different symmetry conserving direction of $A_5$, with,
\begin{equation}
\rho_{\mathbf{3}}(S)\langle \varphi_e \rangle =  \langle \varphi_e \rangle, \qquad
\rho_{\mathbf{3}}(T^3ST^2ST^3S) \langle \varphi_{\mu} \rangle = \langle \varphi_{\mu} \rangle, \qquad
\rho_{\mathbf{3}}(T^3ST^2ST^3)\langle \varphi_{\tau} \rangle = \langle \varphi_{\tau} \rangle\,.
\end{equation}
That is to say
\begin{equation}
\label{eq:G_e_mu_tau}G_{e}=Z^{S}_2,\qquad G_{\mu}=Z^{T^3ST^2ST^3S}_2,\qquad G_{\tau}=Z^{T^3ST^2ST^3}_2\,.
\end{equation}
\begin{table}[t!]
\begin{center}
\begin{tabular}{|c|c|c|c|}\hline\hline
 &  $S$  &   $T$      \\ \hline
~$\mathbf{1}$~ & 1   &  1    \\ \hline
   &   &      \\ [-0.16in]
$\mathbf{3}$ &  $\begin{pmatrix}
 1 &~ 0 ~& 0 \\
 0 &~ -1 ~& 0 \\
 0 &~ 0 ~& -1 \\
\end{pmatrix}$
    & $\frac{1}{2}
\begin{pmatrix}
\phi &~ 1 ~& \phi-1 \\
 -1 &~ \phi-1 ~& \phi \\
 \phi-1 &~ -\phi ~& 1 \\
\end{pmatrix}$\\
&   &     \\ [-0.16in] \hline
&   &      \\ [-0.16in]
$\mathbf{3^\prime}$ &  $\begin{pmatrix}
 1 &~ 0 ~& 0 \\
 0 &~ -1 ~& 0 \\
 0 &~ 0 ~& -1 \\
\end{pmatrix}$
    & $\frac{1}{2}
\begin{pmatrix}
1-\phi &~ \phi ~& 1 \\
 \phi &~ 1 ~& 1-\phi \\
 -1 &~ \phi-1 ~& -\phi \\
\end{pmatrix}$\\
&   &     \\ [-0.16in] \hline
&   &   \\ [-0.16in]
$\mathbf{4}$ & $ \begin{pmatrix}
 -1 &~ 0 ~& 0 ~& 0 \\
 0 &~ -1 ~& 0 ~& 0 \\
 0 &~ 0 ~& 1 ~& 0 \\
 0 &~ 0 ~& 0 ~& 1 \\
\end{pmatrix}$
    & $\frac{1}{4}
\begin{pmatrix}
 -1 &~ -1 &~ 3 &~ \sqrt{5} \\
 1 &~ -3 &~ 1 &~ -\sqrt{5} \\
 -3 &~ 1 &~ 1 &~ -\sqrt{5} \\
 \sqrt{5} &~ \sqrt{5} &~ \sqrt{5} &~ -1 \\
\end{pmatrix} $\\
&   &     \\ [-0.16in] \hline
&   &      \\ [-0.16in]
$\mathbf{5}$ &  $\begin{pmatrix}
 1 &~ 0 &~ 0 &~ 0 &~ 0 \\
 0 &~ 1 &~ 0 &~ 0 &~ 0 \\
 0 &~ 0 &~ -1 &~ 0 &~ 0 \\
 0 &~ 0 &~ 0 &~ -1 &~ 0 \\
 0 &~ 0 &~ 0 &~ 0 &~ 1 \\
\end{pmatrix}$
    & $\frac{1}{8}
\begin{pmatrix}
 1-3 \phi  &~ 2\phi^2 &~ 2/\phi^2  &~ -2 \sqrt{5} &~ \sqrt{3}/\phi \\
 -2\phi^2 &~ -4 &~ 4 &~ 0 &~ -2 \sqrt{3}/\phi  \\
 -2/\phi^2 &~ 4 &~ 0 &~ 4 &~ -2 \sqrt{3}\,\phi  \\
 -2 \sqrt{5} &~ 0 &~ -4 &~ 4 &~ 2 \sqrt{3} \\
 \sqrt{3}/\phi  &~ 2 \sqrt{3}/\phi &~ 2 \sqrt{3}\,\phi  &~ 2 \sqrt{3} &~ 3 \phi -1 \\
\end{pmatrix}$\\
&   &     \\ [-0.16in]\hline\hline
\end{tabular}
\caption{\label{tab:representation} Alternative representation matrices of the generators $S$ and $T$ for the five irreducible representations of $A_5$. This basis is more suitable to discuss the Littlest seesaw model in which the first column of the mixing matrix is in common with $U_{RC}$.}
\end{center}
\end{table}
Inserting these vacuum configurations in Eq.~\eqref{eq:vacuum_clepton} into Eq.~\eqref{eq:Lagrangian_star}, we obtain the charged lepton mass matrix is diagonal with
\begin{equation}
m_{\tau}=y_{\tau}v_{\tau},\quad m_{\mu}=y_{\mu}v_{\mu}, \quad m_{e}=y_{e}v_{e}\,.
\end{equation}
The hierarchies among the three charged lepton masses are expected to be explained by including an extra $U(1)$ symmetry such that the effective Yukawa couplings $y_{\tau}$, $y_{\mu}$ and $y_{e}$ are of different order of magnitudes.

\begin{figure}[t!]
\centering
\begin{tabular}{c}
\includegraphics[width=0.50\linewidth]{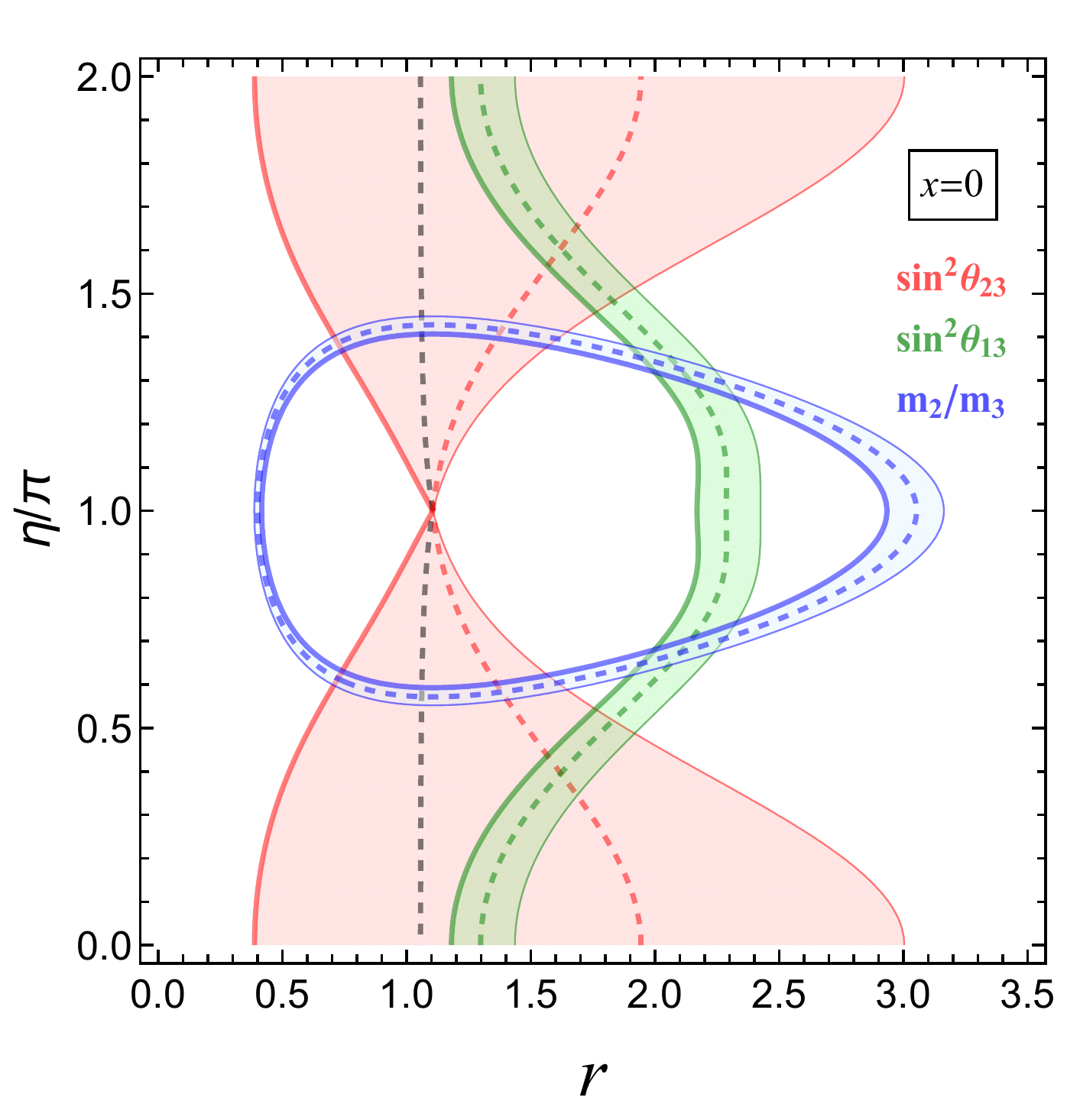}
\end{tabular}
\caption{\label{fig:RC1_LS} Contour plots of $\sin^2\theta_{13}$, $\sin^2\theta_{23}$ and $m_2/m_3$ in the $\eta-r$ plane for the Littlest seesaw model studied in section~\ref{sec:possible_LS}. Here we take $x=0$ for which the solar vacuum alignment $\Phi_{\text{sol}}$ preserves the residual symmetry  $G_{\text{sol}}=Z^{T^2ST}_5$. The $3\sigma$ upper (lower) bounds of the lepton mixing angles are labelled with thick (thin) solid curves, and the dashed contour lines represent the corresponding best fit values. The $3\sigma$ ranges as well as the best fit values of the mixing angles are adapted from~\cite{Capozzi:2017ipn}. The black contour line refers to maximal atmospheric mixing angle with $\sin^2\theta_{23}=0.5$.}
\end{figure}

As regards the neutrino sector, the three columns of the $U_{RC}$ mixing pattern read
\begin{equation}
\Phi_1=\begin{pmatrix}
\phi \\
-1\\
1/\phi
\end{pmatrix},\qquad \Phi_2=\begin{pmatrix}
-1\\
-1/\phi \\
\phi
\end{pmatrix},\qquad \Phi_3=\begin{pmatrix}
1/\phi \\
\phi  \\
1
\end{pmatrix}\,.
\end{equation}
As schematically illustrated in figure~\ref{fig:alignment}, the solar alignment vector $\Phi_{\text{sol}}$ is orthogonal to $\Phi_1$, consequently its most general form is
\begin{equation}
\label{eq:phi_sol}\Phi_{\text{sol}}=\begin{pmatrix}
x  \\
1+x\phi\\
\phi
\end{pmatrix}\,.
\end{equation}
This vacuum alignment $\Phi_{\text{sol}}$ would be enforced by some residual subgroup of $A_5$ for certain value of $x$,
\begin{eqnarray}
\nonumber&& G_{\text{sol}}=Z^{T^2ST}_5,~~~\text{for}~~~x=0,\\
\nonumber&& G_{\text{sol}}=Z^{ST^2ST^3S}_2,~~~\text{for}~~~x=-1,\\
\nonumber&& G_{\text{sol}}=Z^{T^4(ST^2)^2}_2,~~~\text{for}~~~x=1,\\
\nonumber&& G_{\text{sol}}=Z^{(T^2S)^2T^2}_3,~~~\text{for}~~~x=-1/\phi,\\
\label{eq:Phi_sol_symm}&& G_{\text{sol}}=Z^{ST^3ST}_3,~~~\text{for}~~~x=-\phi\,.
\end{eqnarray}
Furthermore the atmospheric alignment vector $\Phi_{\text{atm}}$ is along the direction of $\Phi_2$ or $\Phi_3$ which respects the following residual symmetry
\begin{equation}
G_{\text{atm}}=\left\{\begin{array}{c}
Z^{ST^2ST^3S}_2,~~~\Phi_{\text{atm}}\propto\Phi_2\,,\\
Z^{T^4(ST^2)^2}_2,~~~\Phi_{\text{atm}}\propto\Phi_3\,.
\end{array}
\right.
\end{equation}
In the following we consider the case of $\Phi_{\text{atm}}\propto\Phi_3$~\footnote{The experimental data on mixing angles and neutrino masses can not be accommodated for $\Phi_{\text{atm}}\propto\Phi_2$.}, then the Dirac neutrino mass matrix $M_D$ and the right-handed neutrino mass matrix  $M_{N}$ are
\begin{equation}
M_{D}=\begin{pmatrix}
a/\phi ~&~   x b \\
\phi a  ~&~  (1+\phi x)b \\
a   ~&~  \phi b
\end{pmatrix},\qquad M_{N}=\begin{pmatrix}
M_{\mathrm{atm}}  &  0  \\
0  &   M_{\mathrm{sol}}
\end{pmatrix}\,.
\end{equation}
Applying the seesaw formula results in the effective light neutrino mass matrix
\begin{equation}
\label{eq:mnu_gene_RC1}m_{\nu}=m_a\begin{pmatrix}
2-\phi  ~&  1   ~& \phi-1 \\
1  ~&   \phi+1  ~&  \phi \\
\phi-1  ~&  \phi  ~&  1
\end{pmatrix}+m_{b}e^{i\eta}
\begin{pmatrix}
 x^2 &~ x (x \phi +1) ~& x \phi  \\
 x (x \phi +1) &~ (x \phi +1)^2 ~& \phi  (x \phi +1) \\
 x \phi  &~ \phi  (x \phi +1) ~& \phi +1 \\
\end{pmatrix}\,.
\end{equation}
This neutrino mass matrix $m_{\nu}$ can be simplified into a quite simple form by performing a unitary transformation $U_{RC}$,
\begin{equation}\label{eq:mnu_URC1}
m^{\prime}_{\nu}=U^{T}_{RC}m_{\nu}U_{RC}=\begin{pmatrix}
0  &~   0  &~  0 \\
0  &~  y   &~ z  \\
0  &~  z  &~ w
\end{pmatrix}
\end{equation}
with
\begin{eqnarray}
\nonumber&&y=m_{b}\,e^{i\eta}(x-1)^2,\\
 \nonumber&&z=-m_{b}\,e^{i\eta}\phi\left(x^2-1\right),\\
  &&w=|w|e^{i\phi_{w}}=4m_a+m_{b}\,e^{i\eta}\,\phi^2\left(x+1\right)^2\,.
\end{eqnarray}
The block diagonal neutrino mass matrix $m^{\prime}_{\nu}$ of Eq.~\eqref{eq:mnu_URC1} can be easily diagonalized through the standard procedure, as shown in the appendix~\ref{sec:appendix_B}.
The lepton mixing matrix is predicted to take the form
\begin{equation}
\label{eq:PMNS_URC1}
\hskip-0.1in U=\frac{1}{2}
\begin{pmatrix}
\phi  ~&~ (1-\phi )\sin\theta-e^{i\psi}\cos\theta ~&~ (\phi-1)\cos\theta-e^{i\psi}\sin\theta \\
 -1 ~&~ -\phi\sin\theta+e^{i\psi} (1-\phi )\cos\theta ~&~ \phi \cos
  \theta+e^{i\psi} (1-\phi )\sin\theta \\
\phi-1 ~&~ -\sin\theta+e^{i\psi}\phi\cos\theta ~&~ \cos\theta+e^{i\psi} \phi \sin\theta
\end{pmatrix}P_{\nu}\,,
\end{equation}
with
\begin{equation}
P_{\nu}=\text{diag}(1, e^{i(-\psi+\rho)/2}, e^{i(-\psi+\sigma)/2})\,.
\end{equation}
If we assign the left-handed lepton fields $L=\left(L_{e}, L_{\tau}, L_{\mu}\right)^{T}$ instead of $L=\left(L_{e}, L_{\mu}, L_{\tau}\right)^{T}$ to a triplet of the $A_5$ flavor group, and interchange the vacuum configurations $\langle\varphi_{\mu}\rangle$ and $\langle\varphi_{\tau}\rangle$ in Eq.~\eqref{eq:vacuum_clepton}, the resulting charged lepton mass matrix would be diagonal as well and the lepton mixing matrix can be obtained by exchanging the second and third rows of the PMNS mixing matrix in Eq.~\eqref{eq:PMNS_URC1}. Furthermore, the exact results for the light neutrino masses are given by
\begin{eqnarray}
\nonumber&&m^2_1=0\,,\\
\nonumber&&m^2_2=\frac{1}{2}\left[|y|^2+|w|^2+2|z|^2-\frac{|w|^2-|y|^2}{\cos2\theta}\right]\,,\\
\label{eq:nu_mass_URC1}&&m^2_3=\frac{1}{2}\left[|y|^2+|w|^2+2|z|^2+\frac{|w|^2-|y|^2}{\cos2\theta}\right]\,.
\end{eqnarray}
The expressions for the sine and cosine of rotation angle $\theta$ and the phases $\psi$, $\rho$, $\sigma$ are
\begin{eqnarray}
\nonumber&&\sin2\theta=\frac{2z\,e^{-i\eta}\sqrt{|y|^2+|w|^2+2|y||w|\cos(\phi_{w}-\eta)}}{\sqrt{(|w|^2-|y|^2)^2+4|z|^2\left[|y|^2+|w|^2+2|y||w|\cos(\phi_{w}-\eta)\right]}}\,,\\
\nonumber&&\cos2\theta=\frac{|w|^2-|y|^2}{\sqrt{(|w|^2-|y|^2)^2+4|z|^2\left[|y|^2+|w|^2+2|y||w|\cos(\phi_{w}-\eta)\right]}}\,,\\
\nonumber&&\sin\psi=\frac{|w|\sin(\phi_{w}-\eta)}{\sqrt{|y|^2+|w|^2+2|y||w|\cos(\phi_{w}-\eta)}}\,,\\ \nonumber&&\cos\psi=\frac{|y|+|w|\cos(\phi_{w}-\eta)}{\sqrt{|y|^2+|w|^2+2|y||w|\cos(\phi_{w}-\eta)}}\,,\\
\nonumber&&\sin\rho=-\frac{(m^2_2-|z|^2)\sin\eta+|y||w|\sin\phi_{w}}{m_2\sqrt{|y|^2+|w|^2+2|y||w|\cos(\phi_{w}-\eta)}}\,,\\
\nonumber&&\cos\rho=\frac{(m^2_2-|z|^2)\cos\eta+|y||w|\cos\phi_{w}}{m_2\sqrt{|y|^2+|w|^2+2|y||w|\cos(\phi_{w}-\eta)}}\,,\\
\nonumber&&\sin\sigma=-\frac{(m^2_3-|z|^2)\sin\eta+|y||w|\sin\phi_w}{m_3\sqrt{|y|^2+|w|^2+2|y||w|\cos(\phi_{w}-\eta)}}\,,\\
&&\cos\sigma=\frac{(m^2_3-|z|^2)\cos\eta+|y||w|\cos\phi_w}{m_3\sqrt{|y|^2+|w|^2+2|y||w|\cos(\phi_{w}-\eta)}}\,.
\end{eqnarray}
We can straightforwardly extract the mixing angles from Eq.~\eqref{eq:PMNS_URC1} and find
\begin{eqnarray}
\nonumber&&\sin^2\theta_{13}=\frac{3-\phi}{8}+\frac{1-\phi}{8}\cos2\theta+\frac{1-\phi}{4}\sin2\theta\cos\psi,\\
\nonumber&&\sin^2\theta_{12}=\frac{3-\phi+(\phi-1)\cos2\theta+2(\phi-1)\sin2\theta\cos\psi}{5+\phi+(\phi-1)\cos2\theta+2(\phi-1)\sin2\theta\cos\psi},\\
&&\sin^2\theta_{23}=\frac{3+(2\phi-1)\cos2\theta-2\sin2\theta\cos\psi}{5+\phi+(\phi-1)\cos2\theta+2(\phi-1)\sin2\theta\cos\psi}\,,
\end{eqnarray}
which fulfill the sum rule
\begin{equation}
4\cos^2\theta_{12}\cos^2\theta_{13}=\phi^2\,.
\end{equation}
If inserting the experimental best fit value $\sin^2\theta_{13}=0.0214$~\cite{Capozzi:2017ipn}, we arrive at
\begin{equation}
\sin^2\theta_{12}\simeq0.331\,,
\end{equation}
which is in accordance with the experimental data~\cite{Capozzi:2017ipn}. As regards the Dirac CP phase, we find that the Jarlskog invariant takes a rather simple form,
\begin{equation}
J_{CP}=\frac{1}{16} \sin 2\theta \sin \psi \,,
\end{equation}
and an exact relation for $\cos\delta_{CP}$ in terms of the lepton mixing angles is satisfied,
\begin{equation}
\cos\delta_{CP}=\frac{(\phi-1)\cos^2\theta_{13}+\left[(5+\phi)\sin^2\theta_{13}-3+\phi\right]\cos2\theta_{23}}{2\phi\sqrt{3-\phi-4\sin^2\theta_{13}}\;\sin2\theta_{23}\sin\theta_{13}}\,.
\end{equation}
For the Majorana invariant $I_1$, we get
\begin{eqnarray}
\nonumber I_{1}=&&\frac{2-\phi}{64}  \Big\{4\cos (\rho -\sigma )\left[\cos 2 \theta  \sin 2 \psi- \sin 2 \theta  \sin \psi\right]\\
&&+\sin (\rho -\sigma )\left[ (\cos 4 \theta +3) \cos 2 \psi
-2 \sin 4 \theta  \cos \psi  - \sin ^22 \theta \right]\Big\}\,.
\end{eqnarray}
If $x$ is treated as a free parameter, the experimental data on lepton mixing can be described very well for certain values of $x$, $\eta$
and $r=m_b/m_a$. On the other hand, if we require the solar vacuum alignment is associated with certain residual symmetry, as shown in Eq.~\eqref{eq:Phi_sol_symm}, only $x=0$ is phenomenologically viable. We show how the observables $\sin^2\theta_{13}$, $\sin^2\theta_{23}$ and $m_2/m_3$ vary in the $r-\eta$ plane in figure~\ref{fig:RC1_LS}. In order to show concrete examples, we list the predictions for mixing parameters for some benchmark values of $r$ and $\eta$ in table~\ref{tab:best_fit_URC_LS} and table~\ref{tab:best_fit_URC_LS_23ex}. Note that the atmospheric angle $\theta_{23}$ is outside of $3\sigma$ interval but quite close to $3\sigma$ bounds. We expect this discrepancy could be resolved by considering the contribution of the third almost decoupled right-handed neutrino of mass $M_{\text{dec}}$. Moreover, corrections to the leading order results are generally presented in an explicit model, and therefore it is not difficult to achieve good agreement with experimental data.

\begin{table}[hptb]
\renewcommand{\tabcolsep}{1.7mm}
\centering
\begin{tabular}{| c c c | c c c c c c |}  \hline \hline
\rule{0pt}{2.5ex}%
$\eta$    & $r$ & $x$
  	 & $\sin^2\theta_{13}$  &$\sin^2\theta_{12}$  & $\sin^2\theta_{23}$  & $\delta_{CP}/\pi$ &  $\beta/\pi$  & $m^2_2/m^2_3$   \\ [0.5ex] \hline
\rule{0pt}{2.5ex} $\pm\frac{2 \pi }{3}$ & $2.053$ & $0$ & $0.0223$ & $0.331$ & $0.304$ & $\mp0.424$ & $\pm0.282$ & $0.0289$ \\ [0.5ex] \hline
\rule{0pt}{2.5ex} $\pm\frac{3 \pi }{5}$ & $1.826$ & $0$ & $0.0249$ & $0.329$ & $0.355$ & $\mp0.351$ & $\pm0.200$ & $0.0245$ \\ [0.5ex] \hline
\rule{0pt}{2.5ex} $\pm\frac{5 \pi }{8}$ & $1.904$ & $0$ &  $0.0241$ & $0.329$ & $0.337$ & $\mp0.376$ & $\pm0.228$ & $0.0263$ \\ [0.5ex] \hline
\rule{0pt}{2.5ex} $\pm\frac{7 \pi }{11}$ & $1.942$ & $0$ & $0.0236$ & $0.330$ & $0.328$ & $\mp0.388$ & $\pm0.242$ & $0.0271$ \\ [0.5ex] \hline
\rule{0pt}{2.5ex} $\pm\frac{9 \pi }{14}$ & $1.965$ & $0$ & $0.0234$ & $0.330$ & $0.323$ & $\mp0.396$ & $\pm0.250$ & $0.0275$ \\ [0.5ex] \hline
\rule{0pt}{2.5ex} $\pm\frac{11 \pi }{17}$ & $1.980$ & $0$ &  $0.0232$ & $0.330$ & $0.320$ & $\mp0.400$ & $\pm0.256$ & $0.0278$ \\ [0.5ex] \hline
\rule{0pt}{2.5ex} $\pm\frac{12 \pi }{19}$ & $1.926$ & $0$ &  $0.0238$ & $0.330$ & $0.332$ & $\mp0.383$ & $\pm0.236$ & $0.0268$ \\ [0.5ex] \hline
\rule{0pt}{2.5ex} $\pm\frac{13 \pi }{20}$ & $1.991$ & $0$ &  $0.0231$ & $0.330$ & $0.317$ & $\mp0.404$ & $\pm0.259$ & $0.0280$ \\ [0.5ex] \hline
\rule{0pt}{2.5ex} $\pm\frac{15 \pi }{23}$ & $1.999$ & $0$ &  $0.0230$ & $0.330$ & $0.316$ & $\mp0.406$ & $\pm0.262$ & $0.0281$ \\ [0.5ex] \hline
\rule{0pt}{2.5ex} $\pm\frac{16 \pi }{25}$ & $1.955$ & $0$ & $0.0235$ & $0.330$ & $0.325$ & $\mp0.392$ & $\pm0.246$ & $0.0273$ \\ [0.5ex] \hline
\rule{0pt}{2.5ex} $\pm\frac{17 \pi }{26}$ & $2.005$ & $0$ & $0.0229$ & $0.330$ & $0.314$ & $\mp0.408$ & $\pm0.265$ & $0.0282$ \\ [0.5ex] \hline
\rule{0pt}{2.5ex} $\pm\frac{17 \pi }{27}$ & $1.920$ & $0$ & $0.0239$ & $0.329$ & $0.333$ & $\mp0.381$ & $\pm0.234$ & $0.0266$ \\ [0.5ex] \hline
\rule{0pt}{2.5ex} $\pm\frac{19 \pi }{29}$ & $2.010$ & $0$ & $0.0228$ & $0.330$ & $0.313$ & $\mp0.41$ & $\pm0.266$ & $0.0283$ \\ [0.5ex] \hline
\rule{0pt}{2.5ex} $\pm\frac{19 \pi }{30}$ & $1.932$ & $0$ & $0.0238$ & $0.330$ & $0.330$ & $\mp0.385$ & $0.238$ & $0.0269$ \\ [0.5ex] \hline \hline
\end{tabular}
\caption{\label{tab:best_fit_URC_LS}Benchmark numerical results for the alternative Littlest seesaw model discussed in section~\ref{sec:possible_LS}.}
\end{table}

\begin{table}[hptb]
\renewcommand{\tabcolsep}{1.7mm}
\centering
\begin{tabular}{| c c c | c c c c c c |}  \hline \hline
\rule{0pt}{2.5ex}%
$\eta$    & $r$ & $x$
  	 &  $\sin^2\theta_{13}$  &$\sin^2\theta_{12}$  & $\sin^2\theta_{23}$  & $\delta_{CP}/\pi$ &  $\beta/\pi$  & $m^2_2/m^2_3$   \\ [0.5ex] \hline
\rule{0pt}{2.5ex} $\pm\frac{2 \pi }{3}$ & $2.050$ & $0$ &  $0.0224$ & $0.331$ & $0.696$ & $\pm0.577$ & $\pm0.282$ & $0.029$ \\ [0.5ex] \hline
\rule{0pt}{2.5ex} $\pm\frac{3 \pi }{5}$ & $1.816$ & $0$ &  $0.0251$ & $0.329$ & $0.644$ & $\pm0.65$ & $\pm0.199$ & $0.0246$ \\ [0.5ex] \hline
\rule{0pt}{2.5ex} $\pm\frac{5 \pi }{8}$ & $1.897$ & $0$ &  $0.0243$ & $0.329$ & $0.662$ & $\pm0.625$ & $\pm0.227$ & $0.0264$ \\ [0.5ex] \hline
\rule{0pt}{2.5ex} $\pm\frac{7 \pi }{11}$ & $1.936$ & $0$ & $0.0238$ & $0.33$ & $0.671$ & $\pm0.612$ & $\pm0.241$ & $0.0272$ \\ [0.5ex] \hline
\rule{0pt}{2.5ex} $\pm\frac{9 \pi }{14}$ & $1.959$ & $0$ & $0.0235$ & $0.33$ & $0.676$ & $\pm0.605$ & $\pm0.249$ & $0.0276$ \\ [0.5ex] \hline
\rule{0pt}{2.5ex} $\pm\frac{11 \pi }{17}$ & $1.975$ & $0$ & $0.0233$ & $0.33$ & $0.68$ & $\pm0.6$ & $\pm0.255$ & $0.0279$ \\ [0.5ex] \hline
\rule{0pt}{2.5ex} $\pm\frac{12 \pi }{19}$ & $1.919$ & $0$ &  $0.0240$ & $0.329$ & $0.668$ & $\pm0.618$ & $\pm0.235$ & $0.0269$ \\ [0.5ex] \hline
\rule{0pt}{2.5ex} $\pm\frac{13 \pi }{20}$ & $1.986$ & $0$  & $0.0232$ & $0.33$ & $0.682$ & $\pm0.597$ & $\pm0.259$ & $0.0281$ \\ [0.5ex] \hline
\rule{0pt}{2.5ex} $\pm\frac{15 \pi }{23}$ & $1.994$ & $0$ &  $0.0231$ & $0.33$ & $0.684$ & $\pm0.594$ & $\pm0.262$ & $0.0282$ \\ [0.5ex] \hline
\rule{0pt}{2.5ex} $\pm\frac{16 \pi }{25}$ & $1.949$ & $0$ &  $0.0236$ & $0.33$ & $0.674$ & $\pm0.608$ & $\pm0.245$ & $0.0275$ \\ [0.5ex] \hline
\rule{0pt}{2.5ex} $\pm\frac{17 \pi }{26}$ & $2.000$ & $0$ &  $0.0230$ & $0.33$ & $0.685$ & $\pm0.592$ & $\pm0.264$ & $0.0283$ \\ [0.5ex] \hline
\rule{0pt}{2.5ex} $\pm\frac{17 \pi }{27}$ & $1.912$ & $0$ &  $0.0241$ & $0.329$ & $0.666$ & $\pm0.62$ & $\pm0.232$ & $0.0268$ \\ [0.5ex] \hline
\rule{0pt}{2.5ex} $\pm\frac{19 \pi }{29}$ & $2.005$ & $0$  & $0.0229$ & $0.33$ & $0.686$ & $\pm0.591$ & $\pm0.266$ & $0.0284$ \\ [0.5ex] \hline
\rule{0pt}{2.5ex} $\pm\frac{19 \pi }{30}$ & $1.925$ & $0$ &  $0.0239$ & $0.329$ & $0.669$ & $\pm0.616$ & $\pm0.237$ & $0.027$ \\ [0.5ex] \hline \hline
\end{tabular}
\caption{\label{tab:best_fit_URC_LS_23ex}Benchmark numerical results for the alternative Littlest seesaw model discussed in section~\ref{sec:possible_LS}, where the second and third rows of the mixing matrix in Eq.~\eqref{eq:PMNS_URC1} are exchanged.}
\end{table}

\newpage

\section{\label{sec:introduction}Conclusion}
\indent

The Littlest Seesaw approach assumes that a {\it different} residual flavour symmetry is preserved by each flavon, in the diagonal mass basis of two right-handed neutrinos, leading to a highly predictive set of possible flavon alignments for the charged leptons and neutrinos. The Littlest seesaw model can thereby give a successful description of both neutrino mixing and the light neutrino masses in terms of four input parameters. The case of $S_4$, discussed in earlier work, leads to the lepton mixing matrix being predicted to be of the TM1 form. The neutrino mass spectrum is normal ordered and the lightest neutrino is massless. Moreover, CP violation in neutrino oscillation and leptogenesis arises from a unique single phase such that they are closely related. Therefore the Littlest seesaw model is quite predictive and attractive.

In this work, we have investigated whether the Littlest seesaw is confined to TM1 mixing, or is of more general applicability. We have performed a comprehensive analysis of possible lepton mixing which can be derived from the $A_5$ flavor symmetry group within the paradigm of the Littlest seesaw. The general principle of the Littlest seesaw is that different sectors of the Lagrangian preserve different residual subgroups of the flavor symmetry~\cite{King:2015dvf}. This idea is illustrated in figure~\ref{fig:benz} and figure~\ref{fig:star}. If the residual symmetry of the charged lepton sector is $G_{l}=Z^{T}_5$ which enforces the diagonality of the charged lepton mass matrix in the $T$ generator diagonal basis, the subgroup $G_{\text{atm}}=Z^{T^3ST^2ST^3S}_2$ or $G_{\text{atm}}=Z^{T^3ST^2ST^3}_2$ is preserved by the atmospheric flavon, and solar flavon $\phi_{\text{sol}}$ breaks the flavor group $A_5$ into $G_{\text{sol}}=Z^{T3ST^2S}_3$ or $G_{\text{sol}}=Z^{ST^2ST^3}_3$, the first column of the golden ratio mixing matrix is preserved. The experimental data on the lepton mixing angles and neutrino masses can be accommodated for certain values of the input parameters $m_a$, $m_b$ and $\eta$ except that the reactor angle $\theta_{13}$ is predicted to rather close to its $3\sigma$ boundary. This could be easily reconciled with the experimental results in an explicit model with small subleading corrections or by considering the third almost decoupled right-handed neutrino. Moreover, many numerical benchmark examples are found. The most remarkable point is $\eta=0$ for $G_{\text{atm}}=Z^{T^3ST^2ST^3S}_2$ and $\eta=\pi$ for $G_{\text{atm}}=Z^{T^3ST^2ST^3}_2$, then both Dirac CP phase $\delta_{CP}$ and the atmospheric mixing angle $\theta_{23}$ would be exactly maximal. This mixing pattern is previously predicted in the semidirect approach of  combining $A_5$ flavor symmetry with generalized CP~\cite{Li:2015jxa,DiIura:2015kfa,Ballett:2015wia}, but here we have additional prediction for the neutrino masses and the generalized CP symmetry is not introduced at all.

In the same fashion we find a third golden Littlest seesaw model which preserves the first column of the $U_{RC}$ mixing matrix in Eq.~\eqref{eq:URC}. Accordingly the residual subgroups in different sectors are $G_{e}=Z^{S}_2$, $G_{\mu}=Z^{T^3ST^2ST^3S}_2$, $G_{\tau}=Z^{T^3ST^2ST^3}_2$, $G_{\text{atm}}=Z^{T^4(ST^2)^2}_2$ and $G_{\text{sol}}=Z^{T^2ST}_5$. This case fits the experimental data well to a certain extent. The atmospheric angle $\theta_{23}$ is determined to lie outside the $3\sigma$ region although rather close to $3\sigma$ bounds. Generally corrections to the leading order results are expected to exist in an explicit model such that it is not difficult to achieve agreement with the data. Hence this golden Littlest seesaw model can be regarded as a good leading order approximation from the view of model building.

In conclusion, the Littlest seesaw is a general and predictive framework of explaining neutrino masses and lepton mixing. All the results of this paper only depend on the assumed residual symmetries and they are independent of the underlying mechanism which dynamically realizes the required vacuum alignments. It would be interesting to construct at least one of the above three golden Littlest seesaw models. Since all CP violation phases are completely fixed in the golden Littlest seesaw model, another interesting question is whether the observed baryon asymmetry of the universe can be generated through leptogenesis and the resulting constraints on the right-handed neutrino masses.

\subsection*{Acknowledgements}
G.-J.\, D. and C.-C.\, L. acknowledges the support of the National Natural Science Foundation of China under Grant No 11522546.
S.\,F.\,K. acknowledges the STFC Consolidated Grant ST/L000296/1
and the European Union's Horizon 2020 research and innovation programme under the Marie Sk\l{}odowska-Curie grant agreements
Elusives ITN No.\ 674896 and InvisiblesPlus RISE No.\ 690575. One of the author (G.-J.\, D.) is grateful to Chang-Yuan Yao for his kind help on plotting the figures.

\newpage

\section*{\label{sec:appendix}Appendix}

\begin{appendix}

\section{\label{sec:appendix_A5_group}Group Theory of $A_{5}$}

$A_{5}$ is the group of even permutations of five objects, and it has $5!/2=60$ elements. Geometrically it is the symmetry group of a regular icosahedron.
$A_{5}$ group can be generated by two generators $S$ and $T$ which satisfy the multiplication rules~\cite{Ding:2011cm}:
\begin{equation}
  S^{2}=T^{5}=(ST)^{3}=1\,.
\end{equation}
The 60 element of $A_{5}$ group are divided into 5 conjugacy classes:
\begin{eqnarray}
\nonumber 1C_{1} :&&1\\
\nonumber15C_{2} :&& ST^2ST^3S, TST^4, T^4(ST^2)^2, T^2ST^3, (T^2S)^2T^3S, ST^2ST, S, T^3ST^2ST^3,\\
\nonumber && ~T^3ST^2ST^3S, T^3ST^2, T^4ST^2ST^3S, TST^2S, ST^3ST^2S, T^4ST, (T^2S)^2T^4\\
\nonumber20C_{3} : && ST, TS, ST^4, T^4S, TST^3, T^2ST^2, T^2ST^4, T^3ST, T^3ST^3, T^4ST^2, TST^3S, T^2ST^3S, \\
\nonumber&&~T^3ST^2S, ST^2ST^3, ST^3ST, ST^3ST^2, (T^2S)^2T^2, T^2(T^2S)^2, (ST^2)^2S, (ST^2)^2T^2\\
\nonumber12C_{5}: && T, T^4, ST^2, T^2S, ST^3, T^3S, STS, TST, TST^2, T^2ST, T^3ST^4, T^4ST^3\\
\nonumber12C^{\prime}_5: &&T^2, T^3, ST^2S, ST^3S,(ST^2)^2, (T^2S)^2, (ST^3)^2, (T^3S)^2, (T^2S)^2T^3,\\
&&~T^3(ST^2)^2, T^3ST^2ST^4, T^4ST^2ST^3\,,
\end{eqnarray}
where $nC_k$ denotes a class with $n$ elements which have order $k$.
The group structure of $A_5$ has been exhaustively analyzed in Ref.~\cite{Ding:2011cm}. Following the convention of Ref.~\cite{Ding:2011cm}, we find that $A_{5}$ group has thirty-six abelian subgroups in total: fifteen $Z_{2}$ subgroups, ten $Z_{3}$ subgroups, five $K_{4}$ subgroups and six $Z_{5}$ subgroups.
In terms of the generators $S$ and $T$, the concrete forms of these abelian subgroups are as follows:
\begin{itemize}[leftmargin=1.5em]
\item{$Z_{2}$ subgroups}
\begin{eqnarray}
\nonumber&& Z^{ST^{2}ST^{3}S}_{2}=\{1,ST^{2}ST^{3}S\},\quad  Z^{TST^{4}}_{2}=\{1,TST^{4}\}, \quad Z^{T^{4}(ST^{2})^{2}}_{2}=\{1,T^{4}(ST^{2})^{2}\},\\
\nonumber&& Z^{T^{2}ST^{3}}_{2}=\{1,T^{2}ST^{3}\}, \quad  Z^{(T^{2}S)^{2}T^{3}S}_{2}=\{1,(T^{2}S)^{2}T^{3}S\}, \quad  Z^{ST^{2}ST}_{2}=\{1, ST^{2}ST\},\\
\nonumber&&Z^{S}_{2}=\{1,S\}, \quad   Z^{T^{3}ST^{2}ST^{3}}_{2}=\{1,T^{3}ST^{2}ST^{3}\}, \quad  Z^{T^{3}ST^{2}ST^{3}S}_{2}=\{1,T^{3}ST^{2}ST^{3}S\}, \\
\nonumber && Z^{T^{3}ST^{2}}_{2}=\{1,T^{3}ST^{2}\}, \quad  Z^{T^{4}ST^{2}ST^{3}S}_{2}=\{1,T^{4}ST^{2}ST^{3}S\}, \quad  Z^{TST^{2}S}_{2}=\{1,TST^2S\},\\
\nonumber&& Z^{ST^{3}ST^{2}S}_{2}=\{1, ST^{3}ST^{2}S\}, \quad  Z^{T^{4}ST}_{2}=\{1, T^{4}ST\}, \quad  Z^{(T^{2}S)^{2}T^{4}}_{2}=\{1,(T^{2}S)^{2}T^{4}\}.
\end{eqnarray}
All the above fifteen $Z_{2}$ subgroups are conjugate to each other.
\item{$Z_{3}$ subgroups}
\begin{eqnarray}
\nonumber&&Z^{T^{3}ST^{2}S}_{3}=\{1, T^{3}ST^{2}S,  ST^{3}ST^{2}\},\quad  Z^{TST^{3}S}_3=\{1, TST^{3}S, (ST^{2})^{2}T^{2}\}, \\
\nonumber&& Z^{T^{3}ST}_{3}=\{1, T^{3}ST, T^{4}ST^{2}\}, \quad  Z^{ST}_3=\{1, ST, T^{4}S\}, \\
\nonumber &&  Z^{(T^{2}S)^{2}T^{2}}_{3}=\{1, (T^{2}S)^{2}T^{2}, (ST^{2})^{2}S\}, \quad Z^{TST^{3}}_3=\{1, TST^{3}, T^{2}ST^{4}\},\\
\nonumber&&Z^{T^{2}ST^{2}}_{3}=\{1, T^{2}ST^{2}, T^{3}ST^{3}\}, \quad  Z^{TS}_3=\{1, TS,ST^4\},\\
\nonumber&& Z^{ST^{3}ST}_{3}=\{1, ST^{3}ST, T^{2}(T^{2}S)^{2}\}, \quad  Z^{ST^{2}ST^{3}}_3=\{1, ST^{2}ST^{3}, T^{2}ST^{3}S\}.
\end{eqnarray}
The ten $Z_{3}$ subgroups are related with each other by group conjugation.
\item{$K_{4}$ subgroups}
\begin{eqnarray}
\nonumber && K^{(ST^{2}ST^{3}S, TST^{4})}_{4}\equiv Z^{ST^{2}ST^{3}S}_{2}\times Z^{TST^{4}}_{2}=\{1, ST^{2}ST^{3}S, TST^{4}, T^{4}(ST^{2})^{2}\}, \\
\nonumber && K^{(T^{2}ST^{3}, ST^{2}ST)}_{4}\equiv Z^{T^{2}ST^{3}}_{2}\times Z^{ST^{2}ST}_{2}=\{1, T^{2}ST^{3}, (T^{2}S)^{2}T^{3}S, ST^{2}ST\},\\
\nonumber && K^{(S, T^{3}ST^{2}ST^{3})}_{4}\equiv Z^{S}_{2}\times Z^{T^{3}ST^{2}ST^{3}}_{2}=\{1, S, T^{3}ST^{2}ST^{3}, T^{3}ST^{2}ST^{3}S\}, \\
\nonumber && K^{(T^{3}ST^{2}, TST^{2}S)}_{4}\equiv Z^{T^{3}ST^{2}}_{2}\times Z^{TST^{2}S}_{2}=\{1, T^{3}ST^{2}, T^{4}ST^{2}ST^{3}S, TST^{2}S\},\\
\nonumber && K^{(ST^{3}ST^{2}S, T^{4}ST)}_{4}\equiv Z^{ST^{3}ST^{2}S}_{2}\times Z^{T^{4}ST}_{2}=\{1,ST^{3}ST^{2}S, T^{4}ST, (T^{2}S)^{2}T^{4}\}.
\end{eqnarray}
All the five $K_{4}$ subgroups are conjugate as well.
\item{$Z_{5}$ subgroups}
\begin{eqnarray}
\nonumber &&\hskip-0.35in Z^{STS}_5=\{1, STS, ST^{2}S, ST^{3}S, TST\}, \quad Z^{ST^{3}}_5=\{1,ST^{3}, T^{2}S, (ST^{3})^{2}, (T^{2}S)^{2}\},\\
\nonumber &&\hskip-0.35in Z^{T^{2}ST}_5=\{1, T^{2}ST, T^{4}ST^{3},  T^{3}(ST^{2})^{2},T^{4}ST^{2}ST^{3}\}, \quad  Z^{T}_5=\{1, T, T^{2}, T^{3}, T^{4}\},\\
\nonumber &&\hskip-0.35in Z^{TST^{2}}_5=\{1, TST^{2}, T^{3}ST^{4}, (T^{2}S)^{2}T^{3},T^{3}ST^{2}ST^{4}\},~ Z^{ST^{2}}_5=\{1, ST^{2}, T^{3}S, (ST^{2})^{2}, (T^{3}S)^{2}\}.
\end{eqnarray}
All the six $Z_{5}$ subgroups are related to each other under group conjugation.
\end{itemize}
Here the superscript of a subgroup denotes its generator (or generators). The $A_5$ group has five irreducible representations: one singlet representation $\bf{1}$, two three-dimensional representations $\bf{3}$ and $\bf{3^\prime}$, one four-dimensional representation $\mathbf{4}$ and one five-dimensional representation $\mathbf{5}$.

The character table of $A_{5}$ group is reported in Table \ref{tab:character}. We can straightforwardly obtain the Kronecker products between various representations:
\begin{table}[t]
\begin{center}
\begin{tabular}{|c|c|c|c|c|c|}\hline\hline
 \multirow{2}{*}{$\bf{R}$}  &\multicolumn{5}{c|}{Conjugacy Classes}\\\cline{2-6}
   &$1C_{1}$&$15C_{2}$&$20C_{3}$&$12C_{5}$&$12C^{\prime}_{5}$ \\ \hline
$\bf{1}$&1&1&1&1&1\\\hline

$\bf{3}$&3&$-$1&0& $\phi$ & $1-\phi$
\\\hline

$\bf{3}'$&3&$-1$&0&$1-\phi$& $\phi$ \\\hline

$\bf{4}$&4&0&1&$-1$ &$-1$ \\\hline

$\bf{5}$&5&1&$-1$&0  & 0 \\\hline\hline

\end{tabular}
\caption{\label{tab:character} The character table of the $A_5$ group, where $\phi=\frac{1+\sqrt{5}}{2}$.}
\end{center}
\end{table}
\begin{eqnarray}
\nonumber&&\bf{1}\otimes \bf{R}=\bf{R}\otimes\bf{1}=\bf{R},~~~\bf{3}\otimes\bf{3}=\bf{1}\oplus\bf{3}\oplus\bf{5},~~~\bf{3}'\otimes\bf{3}'=\bf{1}\oplus\bf{3}'\oplus\bf{5}, ~~~\mathbf{3}\times\bf{3}'=\mathbf{4}\oplus\bf{5},\\
\nonumber&&\bf{3}\otimes\bf{4}=\bf{3}'\oplus\bf{4}\oplus\bf{5},~~~\bf{3}'\otimes\bf{4}=\bf{3}\oplus\bf{4}\oplus\bf{5},~~~\bf{3}\otimes\bf{5}
=\bf{3}\oplus\bf{3}'\oplus\bf{4}\oplus\bf{5},\\
\nonumber&&\bf{3}'\otimes\bf{5}=\bf{3}\oplus\bf{3}'\oplus\bf{4}\oplus\bf{5},~~~\bf{4}\otimes\bf{4}=\bf{1}\oplus\bf{3}\oplus\bf{3}'\oplus\bf{4}\oplus\bf{5},
~~~\bf{4}\otimes\bf{5}=\bf{3}\oplus\bf{3}'\oplus\bf{4}\oplus\bf{5_{1}}\oplus\bf{5_{2}},\\ \label{eq:Kronecker}&&\mathbf{5}\otimes\bf{5}=\bf{1}\oplus\bf{3}\oplus\bf{3}'\oplus\bf{4_{1}}\oplus\bf{4_{2}}\oplus\bf{5_{1}}\oplus\bf{5_{2}}.
\end{eqnarray}
where $\bf{R}$ represents any irreducible representation of $A_{5}$, and $\bf{4_{1}}$, $\bf{4_{2}}$, $\bf{5_{1}}$ and $\bf{5_{2}}$ stand for the two $\bf{4}$ and two $\bf{5}$ representations that appear in the Kronecker products.

\section{\label{sec:appendix_B}Diagonalization of a $2\times2$ symmetric
complex matrix }
\cleqn

If neutrinos are Majorana particles, their mass matrix is symmetric and
generally complex. In the following, we present the result for
the diagonalisation of a general $2\times2$ symmetric complex matrix, which
is of the form
\begin{equation}
\mathcal{M}=
\begin{pmatrix}
a_{11}e^{i\phi_{11}}  ~&~  a_{12}e^{i\phi_{12}} \\
a_{12}e^{i\phi_{12}}  ~&~  a_{22}e^{i\phi_{22}}
\end{pmatrix}\,,
\end{equation}
where $a_{ij}$ and $\phi_{ij}$ $(i,j=,1,2)$ are real. It can be diagonalised by a unitary matrix $U$ via
\begin{equation}
U^{T}\mathcal{M}U=\text{diag}(\lambda_1,\lambda_2)\,,
\end{equation}
where the unitary matrix $U$ can be written as
\begin{equation}
U=\begin{pmatrix}
\cos\theta e^{i(\phi+\varrho)/2}   ~&~  \sin\theta e^{i(\phi+\sigma)/2}   \\
-\sin\theta e^{i(-\phi+\varrho)/2}    ~&~    \cos\theta e^{i(-\phi+\sigma)/2}
\end{pmatrix}\,,
\end{equation}
with the rotation angle $\theta$ satisfying
\begin{eqnarray}
&&\tan2\theta=\frac{2a_{12}\sqrt{a^2_{11}+a^2_{22}+2a_{11}a_{22}\cos(\phi_{11}+\phi_{22}-2\phi_{12})}}{a^2_{22}-a^2_{11}}\,.
\end{eqnarray}
The eigenvalues $\lambda_1$ and $\lambda_2$ can always set to be positive
with
\begin{eqnarray}
\nonumber&&\lambda^2_1=\frac{1}{2}\left\{a^2_{11}+a^2_{22}+2a^2_{12}-\mathcal{S}\sqrt{(a^2_{22}-a^2_{11})^2+4a^2_{12}\left[a^2_{11}+a^2_{22}+2a_{11}a_{22}\cos(\phi_{11}+\phi_{22}-2\phi_{12})\right]}\right\}\,,\\
\nonumber&&\lambda^2_2=\frac{1}{2}\left\{a^2_{11}+a^2_{22}+2a^2_{12}+\mathcal{S}\sqrt{(a^2_{22}-a^2_{11})^2+4a^2_{12}\left[a^2_{11}+a^2_{22}+2a_{11}a_{22}\cos(\phi_{11}+\phi_{22}-2\phi_{12})\right]}\right\}\,,
\end{eqnarray}
where
$\mathcal{S}=\text{sign}\big(\left(a^2_{22}-a^2_{11}\right)\cos2\theta\big)$. In the case of $\lambda_{2}>\lambda_{1}$, i.e. $\mathcal{S}=1$, the values of $\sin2\theta$ and $\cos2\theta$ are given by
\begin{eqnarray}
\nonumber \sin2\theta &=& \frac{2a_{12}\sqrt{a^2_{11}+a^2_{22}+2a_{11}a_{22}\cos(\phi_{11}+\phi_{22}-2\phi_{12})}}{\sqrt{(a^2_{22}-a^2_{11})^2+
4a^2_{12}[a^2_{11}+a^2_{22}+2a_{11}a_{22}\cos(\phi_{11}+\phi_{22}-2\phi_{12})]}} \,, \\
\cos2\theta &=& \frac{a^2_{22}-a^2_{11}}{\sqrt{(a^2_{22}-a^2_{11})^2+4a^2_{12}[a^2_{11}+a^2_{22}+2a_{11}a_{22}\cos(\phi_{11}+\phi_{22}-2\phi_{12})]}} \,.
\end{eqnarray}
Finally the phases $\phi$, $\varrho$ and $\sigma$ are given by
\begin{eqnarray}
\nonumber&&\sin\phi=\frac{-a_{11}\sin(\phi_{11}-\phi_{12})+a_{22}\sin(\phi_{22}-\phi_{12})}{\sqrt{a^2_{11}+a^2_{22}+2a_{11}a_{22}\cos(\phi_{11}+\phi_{22}-2\phi_{12})}}=\frac{Im\left(\mathcal{M}_{11}^{*}\mathcal{M}_{12}+\mathcal{M}_{22}\mathcal{M}_{12}^{*}\right)}{\left|\mathcal{M}_{11}^{*}\mathcal{M}_{12}+\mathcal{M}_{22}\mathcal{M}_{12}^{*}\right|},\\
\nonumber&&\cos\phi=\frac{a_{11}\cos(\phi_{11}-\phi_{12})+a_{22}\cos(\phi_{22}-\phi_{12})}{\sqrt{a^2_{11}+a^2_{22}+2a_{11}a_{22}\cos(\phi_{11}+\phi_{22}-2\phi_{12})}}=\frac{Re\left(\mathcal{M}_{11}^{*}\mathcal{M}_{12}+\mathcal{M}_{22}\mathcal{M}_{12}^{*}\right)}{\left|\mathcal{M}_{11}^{*}\mathcal{M}_{12}+\mathcal{M}_{22}\mathcal{M}_{12}^{*}\right|}\,,\\
\nonumber&&\sin\varrho=-\frac{\left(\lambda^2_1-a^2_{12}\right)\sin\phi_{12}+a_{11}a_{22}\sin(\phi_{11}+\phi_{22}-\phi_{12})}{\lambda_1\sqrt{a^2_{11}+a^2_{22}+2a_{11}a_{22}\cos(\phi_{11}+\phi_{22}-2\phi_{12})}}\,,\\
\nonumber&&\cos\varrho=\frac{\left(\lambda^2_1-a^2_{12}\right)\cos\phi_{12}+a_{11}a_{22}\cos(\phi_{11}+\phi_{22}-\phi_{12})}{\lambda_1\sqrt{a^2_{11}+a^2_{22}+2a_{11}a_{22}\cos(\phi_{11}+\phi_{22}-2\phi_{12})}}\,,\\
\nonumber&&\sin\sigma=-\frac{\left(\lambda^2_2-a^2_{12}\right)\sin\phi_{12}+a_{11}a_{22}\sin(\phi_{11}+\phi_{22}-\phi_{12})}{\lambda_2\sqrt{a^2_{11}+a^2_{22}+2a_{11}a_{22}\cos(\phi_{11}+\phi_{22}-2\phi_{12})}}\,,\\
&&\cos\sigma=\frac{\left(\lambda^2_2-a^2_{12}\right)\cos\phi_{12}+a_{11}a_{22}\cos(\phi_{11}+\phi_{22}-\phi_{12})}{\lambda_2\sqrt{a^2_{11}+a^2_{22}+2a_{11}a_{22}\cos(\phi_{11}+\phi_{22}-2\phi_{12})}}\,.
\end{eqnarray}

\end{appendix}

\end{document}